\documentclass[a4paper,fleqn,usenatbib]{mnras}

\usepackage{amsmath,latexsym,mathrsfs}
\usepackage{txfonts}
\usepackage{graphicx,subfigure}
\usepackage{epsfig}
\usepackage{hyperref}
\usepackage{varioref,xr-hyper}
\usepackage[dvipsnames]{xcolor}
\usepackage{multirow}
\usepackage{tikz,array,color,float}
\usepackage[utf8]{inputenc}
\usepackage[T1]{fontenc}

\title[Cosmological direct detection of dark energy]{Cosmological direct detection of dark energy: non-linear structure formation signatures of dark energy scattering with visible matter}

\author[F. Ferlito et al.]{
Fulvio Ferlito,$^{1,2}$\thanks{E-mail: \href{mailto:ferlito@mpa-garching.mpg.de}{ferlito@mpa-garching.mpg.de} (FF)}
Sunny Vagnozzi,$^{3}$\thanks{E-mail: \href{mailto:sunny.vagnozzi@ast.cam.ac.uk}{sunny.vagnozzi@ast.cam.ac.uk} (SV)}\thanks{Newton-Kavli Fellow}
David F. Mota$^{4}$
and Marco Baldi$^{2,5,6}$ \\
$^{1}$Max-Planck-Institut f\"{u}r Astrophysik, Karl-Schwarzschild-Stra\ss e 1, D-85740 Garching bei M\"{u}nchen, Germany\\
$^{2}$Dipartimento di Fisica e Astronomia, Alma Mater Studiorum Universit\`{a} di Bologna, Via Piero Gobetti 93/2, I-40129 Bologna, Italy\\
$^{3}$Kavli Institute for Cosmology, University of Cambridge, Madingley Road, Cambridge CB3 0HA, United Kingdom \\
$^{4}$Institute of Theoretical Astrophysics, University of Oslo, P.O. Box 1029 Blindern, N-0315 Oslo, Norway\\
$^{5}$INAF - Osservatorio di Astrofisica e Scienza dello Spazio di Bologna, Via Piero Gobetti 93/3, I-40129 Bologna, Italy\\
$^{6}$INFN - Sezione di Bologna, viale Berti Pichat 6/2, I-40127 Bologna, Italy}

\date{Accepted XXX. Received YYY; in original form ZZZ}
\pubyear{2022}

\begin{document}

\label{firstpage}

\pagerange{\pageref{firstpage}--\pageref{lastpage}}

\maketitle

\begin{abstract}
We consider the recently proposed possibility that dark energy (DE) and baryons may scatter through a pure momentum exchange process, leaving the background evolution unaffected. Earlier work has shown that, even for barn-scale cross-sections, the imprints of this scattering process on linear cosmological observables is too tiny to be observed. We therefore turn our attention to non-linear scales, and for the first time investigate the signatures of DE-baryon scattering on the non-linear formation of cosmic structures, by running a suite of large N-body simulations. The observables we extract include the non-linear matter power spectrum, halo mass function, and density and baryon fraction profiles of halos. We find that in the non-linear regime the signatures of DE-baryon scattering are significantly larger than their linear counterparts, due to the important role of angular momentum in collapsing structures, and potentially observable. The most promising observables in this sense are the baryon density and baryon fraction profiles of halos, which can potentially be constrained by a combination of kinetic Sunyaev-Zeldovich (SZ), thermal SZ, and weak lensing measurements. Overall, our results indicate that future prospects for cosmological and astrophysical direct detection of non-gravitational signatures of dark energy are extremely bright.
\end{abstract}


\section{Introduction}
\label{sec:intro}

Identifying the nature of dark matter (DM) and dark energy (DE), the two dominant components of the Universe, is among the most pressing issues in physics~\citep[for a review see e.g.][]{Sahni:2004ai}. State-of-the-art strategies for probing the microphysical nature of DM are now focusing on detecting signatures of its \textit{non-gravitational} interactions with visible matter, and include collider~\citep{Boveia:2018yeb}, indirect detection~\citep{Gaskins:2016cha}, and direct detection searches~\citep{MarrodanUndagoitia:2015veg}: in particular, direct detection experiments look for signatures of scattering between DM particles and target nuclei~\citep{Goodman:1984dc}. While not (yet) leading to a convincing detection, these experimental searches for DM have placed strong constraints on its allowed interactions and physical properties, significantly narrowing the space of allowed DM models~\citep{Freese:2017idy}. On the other hand, the microphysical nature of DE is significantly less clear~\citep{Huterer:2017buf}. Some of the most important constraints on DE arise from cosmological measurements of the Universe's geometry and expansion history, which primarily constrain the DE equation of state (EoS) $w_x$, and are consistent with DE being in the form of a cosmological constant (CC) $\Lambda$ with $w_x=-1$~\citep{SupernovaSearchTeam:1998fmf,SupernovaCosmologyProject:1998vns,Pan-STARRS1:2017jku}.

Cosmological background measurements mainly probe DE's \textit{gravitational} interactions, i.e.\ essentially the evolution of the DE energy density as a function of cosmic time. Ultimately, distinguishing between different physical models of DE requires studying the evolution of perturbations~\citep[e.g.][]{Sinha:2021tnr}, through probes such as weak lensing or redshift-space distortions, which have now become mature~\citep{DES:2017qwj,eBOSS:2020yzd,Heymans:2020gsg}. Aside from the possibility of discriminating between different DE models there are other good reasons to move beyond searches for DE's gravitational interactions. From a field theory perspective, if we assume that DE is the manifestation of a new field, which is itself one of the simplest ways of going beyond the CC,~\footnote{The prototypical example of a beyond-$\Lambda$ field model for DE is quintessence, where cosmic acceleration arises as the result of the dynamics of a new ultra-light cosmic scalar field~\citep[see e.g.][]{Wetterich:1987fm,Ratra:1987rm,Caldwell:1997ii,Linder:2007wa,Yang:2018xah}. One concrete construction strongly motivated from particle physics is that where the quintessence field consists of an ultra-light axion or axion-like particle~\citep[see e.g.][]{Hlozek:2014lca,Visinelli:2018utg,Choi:2021aze}. In addition, we further note that the possibility of a negative CC contributing to the DE sector in combination with a quintessence component is in perfect agreement with current cosmological data~\citep[see e.g.][]{Dutta:2018vmq,Visinelli:2019qqu,Calderon:2020hoc,Bonilla:2020wbn,Acquaviva:2021jov,Akarsu:2021fol,Sen:2021wld}. We note that the simplest quintessence models featuring a single scalar field with a canonical kinetic term, minimally coupled to gravity, and without higher derivative operators, appear on their own to be disfavored observationally, as they worsen the $H_0$ tension~\citep{Vagnozzi:2018jhn,Banerjee:2020xcn}.} it is almost impossible to completely decouple DE from other fields, including DM or Standard Model (SM) fields~\citep[see e.g.][]{Wetterich:1987fm,Carroll:1998zi}: even if absent at tree level, couplings of DE to other fields will inevitably arise at loop level, unless protected by a fundamental symmetry. It becomes therefore natural, and one could argue somewhat unavoidable, to consider non-gravitational interactions between DE and other components of the Universe, which in turn might open new ways towards unraveling DE's properties. One important example, covered in a large and growing body of literature, is that of so-called coupled DE models, which involve non-gravitational interactions between DM and DE~\citep{Wang:2016lxa}.~\footnote{For examples of important works on theoretical and observational aspects of coupled DE, see for instance~\cite{Wetterich:1994bg,Amendola:1999er,Mangano:2002gg,Farrar:2003uw,Barrow:2006hia,Pettorino:2008ez,Valiviita:2008iv,Gavela:2009cy,Martinelli:2010rt,DeBernardis:2011iw,Clemson:2011an,Nunes:2014qoa,Benisty:2017eqh,Kumar:2017dnp,Yang:2017ccc,Yang:2018euj,Yang:2018uae,Martinelli:2019dau,Kumar:2019wfs,Yang:2019nhz,Yang:2019uzo,Benetti:2019lxu,DiValentino:2019jae,Lucca:2020zjb,Hogg:2020rdp,Gomez-Valent:2020mqn,Gao:2021xnk,Zhang:2021yof,Benetti:2021div,Kumar:2021eev,Lucca:2021dxo,Lucca:2021eqy}. See also~\cite{Calabrese:2013lga,Euclid:2021cfn} for important studies on DE-electromagnetism couplings.} From the observational perspective, the mismatch between independent measurements of the Hubble constant $H_0$~\citep[the ``Hubble tension'', see][]{Verde:2019ivm,DiValentino:2021izs,DiValentino:2020zio,Perivolaropoulos:2021jda,Shah:2021onj} might be pointing towards a more complex DE sector, potentially featuring non-trivial DE dynamics both at late~\citep{Poulin:2018zxs,Banihashemi:2018oxo,Guo:2018ans,Vagnozzi:2019ezj,Pan:2019hac,Akarsu:2019hmw,Yang:2020zuk,Alestas:2020mvb,Marra:2021fvf,Vagnozzi:2021tjv} and especially early times~\citep{Bernal:2016gxb,Mortsell:2018mfj,Poulin:2018cxd,Knox:2019rjx,Niedermann:2019olb,Sakstein:2019fmf,Ye:2020btb,Zumalacarregui:2020cjh,Hill:2020osr,Ballardini:2020iws,Braglia:2020auw,Vagnozzi:2021gjh,Schoneberg:2021qvd,Ye:2021iwa}.

One feature common to the overwhelming majority of coupled DE models is that they predominantly involve energy exchange between DE and the other component(s) to which DE is coupled (usually DM), with the degree of momentum exchange being small. While these models modify the growth of structure, the energy exchange unavoidably modifies the background expansion as well, which depends no longer solely on the DE EoS $w_x$, but on the dark coupling as well. As a result, models featuring energy exchange are extremely tightly constrained by observational data~\citep[see e.g.][for recent constraints]{Cheng:2019bkh,Pan:2020mst,Yang:2020uga,vonMarttens:2020apn,Nunes:2021zzi}. However, as first pointed out in this context by~\cite{Simpson:2010vh}, low-energy interactions between SM particles typically result in elastic scattering, featuring momentum exchange but negligible energy exchange. In this spirit, \cite{Simpson:2010vh} suggested that DE scattering, i.e.\ DE interactions featuring predominantly momentum rather than energy exchange, might be among the simplest extensions to the simple dark sector picture of the standard $\Lambda$CDM model. These interactions do not modify the background expansion, uniquely determined by $w_x$, but only the evolution of perturbations: this can be compared to Thomson scattering, which only modifies the evolution of the baryon and photon velocity divergences in the cosmological Boltzmann equations~\citep{Ma:1995ey}. Later, \cite{Pourtsidou:2013nha} presented a comprehensive Lagrangian-level taxonomy of coupled DE models~\citep[see also][]{Skordis:2015yra} identifying a class which, within a certain limit, features pure momentum exchange between DE and a second fluid, thus providing a well-motivated theoretical realization of the phenomenological DE scattering class of models of~\cite{Simpson:2010vh}.

Once one accepts the possibility that DE scattering can be the natural outcome of rather minimal theoretical and observational working assumptions, it is natural to consider the possibility that DE might scatter with baryons (visible matter): in analogy to direct detection of DM, this would open up the possibility of \textit{direct detection of DE}.~\footnote{We also note a recent very interesting proposal for the direct detection of DE on Solar System scales put forward in~\cite{He:2017alg} and~\cite{Zhang:2021ygh}, exploiting the gravitational deflection of light. Other proposals for directly search for DE particles can be found for instance in~\cite{Katsuragawa:2016yir} and~\cite{Katsuragawa:2017wge}.} The study of DE-baryon scattering was first concretely pursued by two of us in~\cite{Vagnozzi:2019kvw}, where such an interaction was parametrized in an effective way at the level of the cosmological Boltzmann equations, finding that even for large (barn-scale) values of the DE-baryon scattering cross-section, the imprint on the Cosmic Microwave Background (CMB) and the clustering of the Large-Scale Structure (LSS) is well below the \%-level, and thus unobservably tiny. However, being restricted to linear order in perturbation theory, this earlier study was ill-suited to study the non-linear clustering of the LSS, where the scattering imprints may in principle be large. The goal of our work is to close this gap and extend the study of~\cite{Vagnozzi:2019kvw}, by providing a first investigation of the effects of a non-vanishing DE-baryon scattering cross-section on the non-linear formation of cosmic structures. We run a suite of large N-body simulations incorporating such a scattering, which we find can imprint signatures in the non-linear formation of structures that are significantly larger than their linear counterparts and potentially observable, opening up a possible window towards \textit{cosmological/astrophysical direct detection of DE}.

The rest of this paper is then organized as follows. In Sec.~\ref{sec:scattering} we briefly review how DE-baryon scattering enters the Boltzmann equations and its imprints on cosmological observables in the linear regime (Sec.~\ref{subsec:boltzmann}), we provide an interpretation of this process in terms of an extra force which is useful for our N-body simulation setup (Sec.~\ref{subsec:extraforce}), and finally we discuss how our phenomenological implementation of the DE-baryon scattering process may arise from a well-motivated fundamental Lagrangian (Sec.~\ref{subsec:lagrangian}). In Sec.~\ref{sec:simulations} we explain how the effects of DE-baryon scattering are incorporated in our N-body simulations, and we present our simulation specifications. Sec.~\ref{sec:results} is devoted to discussing our results and is divided into various subsections, each focusing on a specific set of observables extracted from our N-body simulations: general considerations on the large-scale density distribution in Sec.~\ref{subsec:distribution}, non-linear power spectra in Sec.~\ref{subsec:powerspectrum}, halo mass function in Sec.~\ref{subsec:halomassfunction}, halo profiles in Sec.~\ref{subsec:haloprofiles}, and halo baryon fraction profiles in Sec.~\ref{subsec:halobaryonfractionprofiles}. We carry out a critical discussion of our results in Sec.~\ref{sec:discussion}. Finally, in Sec.~\ref{sec:conclusions} we provide concluding remarks, including an outlook on follow-up directions we consider to be of particular interest.

\section{Dark energy-baryon scattering}
\label{sec:scattering}

Here we review the physics of dark energy-baryon scattering. We first discuss how this process affects the Boltzmann equations used to track the evolution of cosmological perturbations in Sec.~\ref{subsec:boltzmann}. In Sec.~\ref{subsec:extraforce} we then discuss an extra force interpretation of this process, useful to set up our N-body simulations. Finally, in Sec.~\ref{subsec:lagrangian} we discuss an underlying Lagrangian realization of the phenomenological model we are studying.

\subsection{Boltzmann equations}
\label{subsec:boltzmann}

We begin by recalling how DE-baryon scattering enters in the system of coupled Boltzmann equations. This discussion will mostly be based on the earlier work of~\cite{Vagnozzi:2019kvw}. We work in synchronous gauge~\citep{Lifshitz:1945du}, a convenient choice as this is the gauge adopted by the Boltzmann solver \texttt{CAMB}~\citep{Lewis:1999bs}. In this gauge, the perturbed Friedmann-Lema\^{i}tre-Robertson-Walker element is given by:
\begin{eqnarray}
ds^2 = a^2(\eta) \left [ -d\eta^2 + (\delta_{ij}+h_{ij})dx^idx^j \right ]\,,
\end{eqnarray}
where $\eta$ denotes conformal time. In Fourier space, we denote the baryon density contrast and velocity divergence by $\delta_b$ and $\theta_b$ respectively, and similarly for the DE density contrast and velocity divergence $\delta_x$ and $\theta_x$. Finally, we denote the DE-baryon slip by $\Theta_{xb} \equiv \theta_x-\theta_b$, and similarly for the photon-baryon slip $\Theta_{\gamma b} \equiv \theta_{\gamma}-\theta_b$, where $\theta_{\gamma}$ is the photon velocity divergence.

Following~\cite{Simpson:2010vh} and~\cite{Vagnozzi:2019kvw}, we assume that DE and baryons scatter elastically, with scattering cross-section given by $\sigma_{xb}$. The cross-section quantifies the likelihood of scattering, and can be viewed as the effective target area seen by the incident particle. In the presence of DE-baryon elastic scattering, the Boltzmann equations for the baryon and DE density contrast and velocity divergence are given by~\citep[see][]{Vagnozzi:2019kvw}:
\begin{eqnarray}
\dot{\delta}_b &=& -\theta_b-\frac{\dot{h}}{2}\,,\label{eq:deltab} \\
\dot{\theta}_b &=& -{\cal H}\theta_b + c_s^2k^2\delta_b + R\tau_c^{-1}\Theta_{\gamma b}+R_x\tau_c^{-1}\alpha_{xb}\Theta_{xb} \,,\label{eq:thetab} \\
\dot{\delta}_x &=& -(1+w_x) \left ( \theta_x+\frac{\dot{h}}{2} \right )-3{\cal H}(c_{s,x}^2-w_x)\delta_x \nonumber \\
&& -9{\cal H}^2(c_{s,x}^2-w_x)(1+w_x)\frac{\theta_x}{k^2} \,, \label{eq:deltade} \\
\dot{\theta}_x &=& -{\cal H}(1-3c_{s,x}^2)\theta_x+\frac{c_{s,x}^2k^2}{1+w_x}\delta_x-\tau_c^{-1}\alpha_{xb}\Theta_{xb} \,. \label{eq:thetade}
\end{eqnarray}
In the above Eqs.~(\ref{eq:deltab}--\ref{eq:thetade}), $h$ is the usual synchronous gauge metric perturbation~\citep[see][]{Ma:1995ey}, ${\cal H}$ is the conformal Hubble parameter, $c_s^2$ is the baryon sound speed squared, and $c_{s,x}^2$ is the DE sound speed squared, which throughout this work we will fix to $c_{s,x}^2=1$. Moreover, we have assumed that the DE EoS $w_x$ is constant, so that the adiabatic DE sound speed squared is given by $c_{a,x}^2=w_x$. The photon-to-baryon and DE-to-baryon density ratios are given by $R \equiv 4\rho_{\gamma}/3\rho_b$ and $R_x \equiv (1+w_x)\rho_x/\rho_b$ respectively, with $\rho_b$, $\rho_{\gamma}$, and $\rho_x$ being the baryon, photon, and DE energy densities. The Thomson scattering opacity is given by $\tau_c \equiv (an_e\sigma_T)^{-1}$, where $a$ is the scale factor, $n_e$ is the number density of electrons, and $\sigma_T \approx 6.7 \times 10^{-25}\,{\rm cm}^2 = 0.67\,{\rm b}$ is the value of the Thomson scattering cross-section which quantifies the strength of baryon-photon scattering (with $1\,{\rm b}=10^{-24}\,{\rm cm}^2$ defining the barn unit). Finally, the ``Thomson ratio'' $\alpha_{xb}$ is given by $\alpha_{xb} \equiv \sigma_{xb}/\sigma_T$, i.e.\ it is numerically equivalent to the DE-baryon scattering cross-section in units of the Thomson scattering cross-section.

A note is in order regarding our assumptions on $c_{s,x}^2$ and $c_{a,x}^2$, and the fact that the two are assumed to be different in Eqs.~(\ref{eq:deltab}--\ref{eq:thetade}). Recall that the sound speed squared $c_{s,x}^2$ is defined as the ratio between DE pressure and density perturbations in the DE rest frame:
\begin{eqnarray}
c_{s,x}^2 = \frac{\delta P_x}{\delta \rho_x}\Bigg\vert_{\rm rf}\,.
\label{eq:csx2}
\end{eqnarray}
In a general frame, $\delta P_x$ and $\delta \rho_x$ are instead related by:
\begin{eqnarray}
\delta P_x = c_{s,x}^2\delta\rho_x + 3{\cal H} \left ( 1+w_x \right ) \left ( c_{s,x}^2-c_{a,x}^2 \right ) \frac{\theta_x\rho_x}{k^2}\,,
\label{eq:pressuredensityperturbationsgeneralframe}
\end{eqnarray}
where the DE adiabatic sound speed squared $c_{a,x}^2$ is given by the ratio between the time derivatives of the DE background pressure and energy density:
\begin{eqnarray}
c_{a,x}^2 \equiv \frac{\dot{P}_x}{\dot{\rho}_x}\,.
\label{eq:cax2}
\end{eqnarray}
For a non-interacting DE fluid, Eq.~(\ref{eq:cax2}) reduces to:
\begin{eqnarray}
c_{a,x}^2 = w_x-\frac{\dot{w}_x}{3{\cal H}(1+w_x)}\,.
\label{eq:cax2new}
\end{eqnarray}
For models of DE based on a single light, minimally coupled scalar field, with a canonical kinetic term, and in the absence of higher order operators (as in standard quintessence models), we expect $c_{s,x}^2 \approx 1$. If, in addition, the DE EoS is a constant as we have assumed in this work, $\dot{w}_x=0$ in Eq.~(\ref{eq:cax2new}) which therefore simplifies to $c_{a,x}^2=w_x$. Therefore, our assumption of setting $1 = c_{s,x}^2 \neq c_{a,x}^2=w_x$ is not unrealistic~\citep[see e.g.][for further discussions]{Hu:1998kj,Bean:2003fb,Ballesteros:2010ks}.

In the presence of DE interactions, the denominator of the right-hand side of Eq.~(\ref{eq:cax2new}) receives a correction proportional to the background energy exchange rate. However, in our phenomenological model there is no background energy exchange up to linear order (see also the later discussion in Sec.~\ref{subsec:lagrangian}), so that we can still safely assume that $c_{a,x}^2$ is given by Eq.~(\ref{eq:cax2new}), and therefore $c_{a,x}^2=w_x$ for a constant DE EoS. A similar point had been noted earlier by one of us in~\cite{Asghari:2019qld} in the context of a DM-DE elastic scattering model, which therefore shares similar features to ours (including the absence of energy exchange at the background level).

We note that in principle we can write Eqs.~(\ref{eq:thetab},\ref{eq:thetade}) in terms of the DE-baryon scattering opacity $\tau_x \equiv (an_e\sigma_{xb})^{-1} = \tau_c\sigma_T/\sigma_{xb} = \tau_c/\alpha_{xb}$. However, for numerical purposes and for ease of comparison to the earlier work of~\cite{Vagnozzi:2019kvw}, it is more convenient to work with the dimensionless Thomson ratio. Moreover, we note that the appearance of the factor $(1+w_x)$ in the DE-to-baryon density ratio $R_x$ is required to enforce total momentum conservation, which in turn follows from the elastic nature of the scattering process we are considering. The term proportional to the DE-baryon slip $\Theta_{xb}$ in Eqs.~(\ref{eq:thetab},\ref{eq:thetade}), depending on the sign of $(1+w_x)$ and therefore on whether the DE component lies in the \textit{quintessence}-like $(w_x>-1)$ or \textit{phantom} $(w_x<-1)$ regime, corresponds respectively to a friction or drag term for the baryon velocity field. In fact, as we will discuss later, our implementation of the scattering in our N-body simulations can indeed be readily interpreted as the effect of an extra friction or drag force felt by baryonic particles. We further note that $\vert n_e\sigma_{xb}\Theta_{xb} \vert$ gives the fraction of DE quanta scattered off baryons per unit time~\citep{Vagnozzi:2019kvw}.

Two clarifications are in order. First of all, we point out that as there is only momentum and no energy transfer between DE and baryons, the background evolution is unchanged by the presence of the DE-baryon scattering (analogously to how Thomson scattering does not alter the background expansion in the early Universe). In other words, the continuity equations for the baryon and DE energy densities are unaffected, and the background expansion is identical to that of a reference no-scattering $w$CDM cosmological with the same DE EoS $w_x \neq -1$, but where $\alpha_{xb}=0$ so that DE and baryons do not scatter. Furthermore, we note that the presence of the $(1+w_x)$ factor in the DE-to-baryon density ratio $R_x$ (related to momentum conservation) implies that scattering can only take place if the DE EoS is $w_x \neq -1$, i.e.\ if DE is not in the form of a CC. The reason why a CC cannot scatter is related to its being smooth and not featuring perturbations $(\delta_x=\theta_x=0)$, as already noted earlier in~\cite{Simpson:2010vh} and~\cite{Vagnozzi:2019kvw}.

We also comment on related cosmological scenarios explored in the literature. After~\cite{Vagnozzi:2019kvw}, also~\cite{Jimenez:2020ysu} studied the possibility of DE-baryon elastic scattering. However, unlike the case studied in~\cite{Vagnozzi:2019kvw} and considered here, \cite{Jimenez:2020ysu} envisaged a scenario where the DE-baryon coupling is time-dependent, which was argued to be a potentially natural scenario if DE is the manifestation of a scalar field $\phi$ and the scattering is mediated by the gradient of the field itself. Here, in the interest of simplicity and since we are not committing to any fundamental model~\citep[as well as for ease of comparison to the results of][]{Vagnozzi:2019kvw}, we shall keep investigating the constant coupling scenario, but note that it would certainly be interesting to further study cases where the DE-baryon coupling is time-dependent, particularly as this might enhance the detection prospects of the DE-baryon scattering signatures. We defer a full study of this well-motivated scenario to follow-up work.

In closing, we also note that cosmological scenarios featuring elastic scattering between various cosmic components (usually one of these being either DM or DE) have been studied in a large body of literature. For example, the closely related case of DE-DM elastic scattering has been covered in various works, which besides the previously discussed~\cite{Simpson:2010vh} include for instance~\cite{Xu:2011nr,Richarte:2014yva,Boehmer:2015sha,Tamanini:2015iia,Koivisto:2015qua,Pourtsidou:2016ico,Dutta:2017kch,Kumar:2017bpv,Linton:2017ged,Bose:2017jjx,Bose:2018zpk,Asghari:2019qld,Kase:2019veo,Kase:2019mox,Chamings:2019kcl,Asghari:2020ffe,Amendola:2020ldb,DeFelice:2020icf,BeltranJimenez:2020qdu,Figueruelo:2021elm,Jimenez:2021ybe,Carrilho:2021hly,Linton:2021cgd,Mancini:2021lec,Carrilho:2021rqo} among others. Other studies in the literature have instead investigated cosmological signatures of different types of (elastic) scattering involving DM, including DM-photon scattering~\citep{Wilkinson:2013kia,Stadler:2018jin,Kumar:2018yhh,Yadav:2019jio}, DM-neutrino scattering~\citep{Serra:2009uu,Wilkinson:2014ksa,Escudero:2015yka,DiValentino:2017oaw,Stadler:2019dii,Hooper:2021rjc}, DM-baryon scattering~\citep{Dvorkin:2013cea,Gluscevic:2017ywp,Boddy:2018kfv,Xu:2018efh,Boddy:2018wzy,Ali-Haimoud:2021lka,Buen-Abad:2021mvc,Nguyen:2021cnb,Rogers:2021byl}, DM self-scattering or scattering with a dark radiation component~\citep{Cyr-Racine:2015ihg,Vogelsberger:2015gpr,Archidiacono:2017slj,Buen-Abad:2017gxg,Archidiacono:2019wdp}, as well as ``multi-interacting DM'' scenarios featuring multiple such interactions simultaneously~\citep{Becker:2020hzj}.

In~\cite{Vagnozzi:2019kvw}, some of us investigated the (linear) cosmological implications of DE-baryon scattering as described by Eqs.~(\ref{eq:deltab}--\ref{eq:thetade}). There it was found that even for rather large values of the Thomson ratio $\alpha_{xb} \approx {\cal O}(1)$, the imprint of DE-baryon scattering on linear cosmological observables is unobservably small. This is true for realistic values of the DE EoS $w_x$ (i.e.\ values not too far from the CC case $w_x=-1$), a clarification which is important as the value of $\vert 1+w_x \vert$ parametrically controls the strength of the DE-baryon scattering effects. For a quintessence-like DE component, i.e.\ one where $w_x>-1$, the effect of the scattering is to decrease the amplitude of DE perturbations, in turn easing the decay of gravitational potentials~\citep{Weller:2003hw,Calabrese:2010uf}. The reverse occurs for a phantom DE component, i.e.\ one where $w_x<-1$, although the physical interpretation of phantom DE models is generally not simple~\citep[see e.g.][for examples of fundamental physics realizations of effective phantom scenarios]{Caldwell:1999ew,Caldwell:2003vq,Vikman:2004dc,Carroll:2004hc,Deffayet:2010qz,Nunes:2015rea,Cognola:2016gjy}.

Considering a quintessence-like DE component, at the level of the CMB the increased decay of gravitational potentials translates into an enhanced late integrated Sachs-Wolfe (LISW) effect~\citep{Sachs:1967er}, leading to increased temperature anisotropy power on large angular scales (small multipoles $\ell$). However, even for $\alpha_{xb} \sim {\cal O}(1)$, the increase in power is at most at the ${\cal O}(1)\%$ level, well below the cosmic variance level and hence totally unobservable~\citep{Vagnozzi:2019kvw}. For a phantom DE component and at a fixed value of $\vert 1+w_x \vert$, the effects are reversed in sign, i.e.\ the LISW amplitude is suppressed, but remain of comparable magnitude and hence unobservable~\citep{Vagnozzi:2019kvw}.

For what concerns the linear matter power spectrum, i.e.\ for sufficiently small wavenumber $k$, the impact of DE-baryon scattering was found to be even smaller than for the case of the CMB. As already anticipated, for a quintessence-like DE component the effect of the scattering effectively corresponds to an extra friction force felt by the baryons. This friction slows down the growth of structure, hence suppressing LSS clustering (and thus the matter power spectrum) on large scales, where the suppression is to very good approximation scale-independent and hence degenerate with a decrease in $\sigma_8$~\citep{Vagnozzi:2019kvw}. The effects are again reversed in sign for a phantom DE component, in which case the scattering leads to an extra drag force. However, on linear scales all these effects were found to be at the ${\cal O}(0.1)\%$ level or smaller, and hence unobservably small~\citep{Vagnozzi:2019kvw}.

The main caveat to all the results discussed above and reported in~\cite{Vagnozzi:2019kvw} is that these were obtained working at linear order in perturbation theory. The latter is sufficient to study the impact of DE-baryon scattering on the CMB and the linear matter power spectrum, but not on the non-linear or mildly non-linear matter power spectrum. There is, however, good reason to believe that DE-baryon scattering may leave interesting signatures in the non-linear matter power spectrum, and more generally in the non-linear formation of structures, signatures which could be conceivably much larger than their linear counterparts and therefore potentially detectable. To understand why, it is useful to consider the related case of DM-DE elastic scattering, whose linear cosmological effects were studied in~\cite{Simpson:2010vh}. The non-linear signatures of the same model were studied by one of us in~\cite{Baldi:2014ica,Baldi:2016zom} by running a suite of N-body simulations, which showed that deviations in the non-linear matter power spectrum compared to the no-scattering case were not only reversed in sign, but had an amplitude larger by more than an order of magnitude compared to their linear counterparts.

We can reasonably expect that the same could potentially occur for the case of DE-baryon scattering. Therefore, our goal in this paper is to extend the earlier work of~\cite{Vagnozzi:2019kvw} by running an appropriate set of N-body simulations to study the effect of DE-baryon scattering on the non-linear formation of cosmic structures, and investigate whether the earlier conclusion concerning the undetectability of the effects of this scattering persists even at non-linear level. It is worth pointing out that there is a large body of literature concerning N-body simulations of non-standard DE models, several of which have focused on coupled DE models featuring energy exchange between DM and DE~\citep[see e.g.][]{Maccio:2003yk,Baldi:2008ay,Li:2010re,Baldi:2010td,Baldi:2010ks,Li:2010zzx,Baldi:2011wa,Baldi:2011qi,Marulli:2011jk,Beynon:2011hw,Cui:2012is,Giocoli:2013ba,Carbone:2013dna,Moresco:2013nfa,Carlesi:2014kua,Carlesi:2014faa,Sutter:2014gna,Pace:2014tya,Maccio:2015iya,Bonometto:2015mya,Penzo:2015tha,Casas:2015qpa,Giocoli:2018ens,Hashim:2018dek,Zhang:2018glx}. To the best of our knowledge, N-body simulations of momentum exchange between DE and DM were performed only in~\cite{Baldi:2014ica,Baldi:2016zom} by one of us, whereas the present work is the first to explore momentum exchange between DE and baryons.

\subsection{Extra force interpretation}
\label{subsec:extraforce}

In order to later set up our N-body simulations, it is useful to phrase the effect of DE-baryon scattering in terms of an additional 4-force $f^{\mu}$ felt by baryonic particles, following earlier discussions in~\cite{Baldi:2014ica,Baldi:2016zom}. Consider a particle with 4-velocity $u^{\mu}=(\gamma,\gamma\mathbf{v})$, where $\gamma$ is the Lorentz factor and $\mathbf{v}$ is the velocity 3-vector with norm $v$. Now consider the scenario where this particle crosses an isotropic perfect fluid with EoS $w$, whose stress-energy tensor is therefore given by $T_{\mu\nu}={\rm diag}(\rho,w\rho,w\rho,w\rho)$. In this case it can be shown that, as long as $w \neq -1$, the particle observes a non-zero momentum flux $\Bar{T}_0^{\mu}$. The result of this is that the particle feels an extra 4-force $f^{\mu}$ proportional to the scattering cross-section $\sigma$. As shown in~\cite{Padmanabhan:1994nc}, $f^{\mu}$ is given by:
\begin{eqnarray}
f^{\mu} = \sigma \left [ T^{\mu}_{\nu}u^{\nu} - u^{\mu} \left ( T_{\rho\sigma}u^{\rho}u^{\sigma} \right )  \right ]\,,
\label{eq:fmu}
\end{eqnarray}
from which it can be trivially shown that $g^{\mu}u_{\mu}=0$, in other words that the extra force is orthogonal to the particle's 4-velocity. Expanding the right-hand side of Eq.~(\ref{eq:fmu}), the extra 4-force can be rewritten in a convenient form:
\begin{eqnarray}
f^{\mu} = (\gamma\mathbf{F} \cdot \mathbf{v},\gamma\mathbf{F})\,,
\label{eq:fmuexpanded}
\end{eqnarray}
where the force 3-vector $\mathbf{F}$ is given by:
\begin{eqnarray}
\mathbf{F} = -(1+w)\sigma\gamma^2\rho\mathbf{v}\,.
\label{eq:f}
\end{eqnarray}
One well-known example of force as in Eq.~(\ref{eq:f}) is relevant for early-Universe cosmology, where scattering between the background radiation with EoS $w=1/3$ and electrons, with cross-section $\sigma_T$, leads to a drag force~\citep{Eisenstein:1997ik} with magnitude:
\begin{eqnarray}
F = -\frac{4}{3}\sigma_Tv\rho_{\gamma}\,.
\label{eq:thomson}
\end{eqnarray}
The length of the sound horizon at the epoch when electrons are released from this drag force is imprinted in the clustering of the LSS, and can serve as a standard ruler to measure distances at low redshift~\citep{Eisenstein:1997ik,Eisenstein:1998tu,SDSS:2005xqv}.

As in the Boltzmann equations Eqs.~(\ref{eq:deltab}--\ref{eq:thetade}), the appearance of the factor $(1+w)$ in Eq.~(\ref{eq:f}) makes it clear that scattering can only occur with a component whose EoS is $w \neq -1$, else the two contributions to Eq.~(\ref{eq:fmu}) cancel, leading to no extra force. It is also worth commenting on the sign of the extra force in Eq.~(\ref{eq:f}). For a quintessence-like DE component, $\mathbf{F}$ is anti-aligned with respect to $\mathbf{v}$, and therefore Eq.~(\ref{eq:f}) effectively describes a \textit{friction} force. The reverse occurs for a phantom DE component, which instead corresponds to a \textit{drag} force. This extra force interpretation also helps clarify our earlier discussion on why scattering with a quintessence-like [phantom] DE component suppresses [enhances] the matter power spectrum on linear scales.

\subsection{Underlying Lagrangian}
\label{subsec:lagrangian}

So far we have introduced the DE-baryon elastic scattering interaction at a purely phenomenological level. While this suffices for the purposes of the present work, it is nonetheless desirable to investigate whether such a phenomenological behavior can arise from a well-motivated fundamental Lagrangian. If we assume that DE is described by a scalar field $\phi$, pure momentum exchange couplings can arise from couplings of the covariant derivative of $\phi$ to the velocity field of the scattering species. Within the three-fold classification of coupled DE models presented in~\cite{Pourtsidou:2013nha,Skordis:2015yra}, this would correspond to so-called \textit{Type 3} models, although in our case with the DM velocity field replaced by the baryon velocity field.

If we denote the baryon velocity 4-vector by $u_b^{\mu}$, the underlying Lagrangian is given by the following~\citep{Pourtsidou:2013nha}:
\begin{eqnarray}
{\cal L} \supset F(\nabla_{\mu}\phi\nabla^{\mu}\phi,u_b^{\mu}\nabla_{\mu}\phi,\phi)+f(n_b)\,,
\label{eq:type3}
\end{eqnarray}
where $F$ and $f$ are in principle arbitrary functions, and $n_b$ is the baryon fluid number density, although the $f(n_b)$ term can be discarded for the purposes of the following discussion. It is precisely the vector interaction structure in the above Lagrangian which allows for the extra force to be either attractive or repulsive (friction or drag), as previously discussed in Sec.~\ref{subsec:extraforce}. If we define $Z \equiv u_b^{\mu}\nabla_{\mu}\phi$, $F_Z \equiv dF(Z)/dZ$, and $q_{\mu}^{\nu}=u_{\mu}u^{\nu}+\delta_{\mu}^{\nu}$, the coupling current vector $J^{\mu}$ is then given by~\citep{Pourtsidou:2013nha}:
\begin{eqnarray}
J^{\mu} = q_{\alpha}^{\mu} \left [ \nabla_{\nu} \left ( F_Zu^{\nu} \right ) \nabla^{\alpha}\phi + F_Z\nabla^{\alpha}Z + ZF_Zu^{\nu}\nabla_{\nu}u_{b}^{\alpha} \right ] \,.
\label{eq:jmu}
\end{eqnarray}
From the above, it can be shown that in the DE rest frame the time component of $J^{\mu}$, which is related to the energy transfer, is $J_0=0$ up to second order. On the other hand, the spatial components satisfy $J_i \neq 0$, so that the Lagrangian in Eq.~(\ref{eq:type3}) describes pure momentum exchange (at least) up to linear order.

\cite{Skordis:2015yra} argued that a formal equivalence between Type 3 theories and phenomenological models of elastic dark scattering cannot be established. Heuristically, the reason is that within the phenomenological dark elastic scattering model the interaction is tied to the local energy density, whereas in Type 3 theories it depends on the gradient of the scalar field (which is of course not a localized quantity). Nonetheless, it can be shown~\citep[see e.g.][]{Skordis:2015yra,Baldi:2016zom} that within the regime $\vert d\ln F(Z)/dZ \vert \gg 1$ (which can be realized for example by the choice $F(Z) \propto e^{-Z}$) the two scenarios in question match to very good approximation, particularly when considering the precision of cosmological observables. In this case, and to leading order in $Z$, the effective cross-section $\sigma_{\rm eff}$ which emerges from the Type 3 Lagrangian is given by~\citep[see e.g.][]{Baldi:2016zom}:
\begin{eqnarray}
\sigma_{\rm eff} \simeq -\frac{3{\cal H}a^2F_Z}{n_{b,0}Z}\,,
\label{eq:sigmaeff}
\end{eqnarray}
where $n_b = n_{b,0}a^{-3}$ is the proper number density of baryons.

Finally, one finds that within this regime of Type 3 theories and on sub-horizon scales, the momentum flux is aligned with the velocity divergence field of the scattering species~\citep{Baldi:2016zom}. At the microphysical level, this implies that the scattering species (in this case baryons) will be subject to an extra force which is aligned or anti-aligned with their velocity vector (a drag or friction force respectively), where the distinction between the two cases depends once more on the sign of $(1+w_x)$. This agrees with the extra force interpretation presented in Sec.~\ref{subsec:extraforce}, and which is at the heart of the implementation of our N-body simulations.

In closing, we also note that one should be very careful when extrapolating the linear theory behavior captured by Eqs.~(\ref{eq:deltab}--\ref{eq:thetade}) down to local scales, for instance when discussing the possible complementarity with terrestrial experiments. In general, interactions between DE and baryonic matter will need to be appropriately screened on local scales, in order to pass local gravity tests~\citep{Carroll:1998zi}. Therefore, a mapping between cosmological and local scales can strictly speaking only be established in the context of an UV complete theory which encompasses the scattering process (here introduced at a purely phenomenological level) and is equipped with a screening mechanism. One example in this sense is the scenario studied in~\cite{Vagnozzi:2021quy}, which envisages scattering between baryons and a DE component equipped with a screening mechanism of the chameleon type~\citep[see][]{Khoury:2003rn,Khoury:2003aq,Mota:2006ed,Burrage:2016bwy,Burrage:2017qrf,Sakstein:2018fwz,Cai:2021wgv}.

The question of how to incorporate a realistic screening mechanism in our simulations is a highly non-trivial one. Ultimately, addressing this requires committing to a fundamental theoretically motivated framework incorporating the screening mechanism, thus moving beyond the phenomenological picture we are considering. In several screening mechanisms, whether or not a region is screened depends on the local value of a given quantity with respect to a critical value of the same quantity (set by the underlying theory): in a number of well-motivated screening models, such quantity could be the local gravitational potential, gravitational force, or force gradient~\citep[for a recent review on the topic see for instance][]{Brax:2021wcv}.

A phenomenological approach towards incorporating screening effects in our simulations could consist in combining our unscreened dynamics with a phenomenological ``screening factor'', which would more or less smoothly interpolate between the unscreened and screened regimes, depending on the local value of the ratio of a given relevant quantity with respect to a reference value: an example of such a quantity could be the local gravitational potential, which would be the relevant quantity for the chameleon~\citep{Khoury:2003rn} and symmetron~\citep{Hinterbichler:2010es} mechanisms, or in alternative a readily available handy proxy for the local gravitational potential such as the Smoothed Particle Hydrodynamics (SPH) density.~\footnote{We note that a method to simulate the non-linear dynamics of screening within a large number of modified gravity theories by combining the linear Klein-Gordon equation with a screening factor has been presented in~\cite{Winther:2014cia}.} The introduction of screening mechanisms could potentially lower all the effects we will discuss later, although the full extent of this will depend sensitively on the screening mechanism adopted. For instance, constraints on chameleon- and symmetron-screened fifth forces on galaxy and cluster scales are still rather permissive, allowing for relative deviations in Newton's constant of up to order $10\%$ or larger~\citep[see e.g.][]{Cataneo:2014kaa,Desmond:2018euk,Desmond:2018sdy,Desmond:2018kdn}, much weaker than limits on local scales. It is therefore possible that these types of screening mechanisms will not dramatically reduce our predictions. Nonetheless, a complete study of the impact of introducing screening mechanisms on our results is beyond the scope of the present work, and will be examined in planned follow-up work focused on running more realistic simulations.

\section{N-body simulations}
\label{sec:simulations}

\begin{table}
\centering
\begin{tabular}{c c c} 
\hline \hline \\[-1em]
Simulation & $w_x$ & $\alpha_{xb}$  \\ [0.5ex] \hline\hline \\[-1em]
 $\;\;\; \Lambda$CDM$\;\;\;$ & $-1$ & $-$ \\ \hline \\[-1em]
Q0  & $-0.9$ & 0 \\ \hline \\[-1em]
Q1 & $\;\;-0.9\;\;$ & 1 \\ \hline \\[-1em]
Q10  & $-0.9$ & 10 \\ \hline \\[-1em]
Q100 & $-0.9$ & $\;\; 100 \;\;$ \\ \hline \\[-1em]
P0  & $-1.1$ & 0 \\ \hline \\[-1em]
P1  & $-1.1$ & 1   \\ \hline \\[-1em]
P10  & $-1.1$ & 10 \\ \hline \\[-1em]
P100 & $-1.1$ & 100 \\ \hline \hline
\end{tabular}
\caption{Suite of N-body simulations performed with our modified version of \texttt{GADGET-3}. Every simulation is characterized by the DE equation of state $w_{x}$ and the DE-baryon scattering cross-section $\sigma_{xb}$ in terms of the Thomson ratio $\alpha_{xb}\equiv\sigma_{xb}/\sigma_T$, where $\sigma_T \approx 6.7 \times 10^{-25}\,{\rm cm}^2 = 0.67\,{\rm b}$ is the Thomson scattering cross-section. The nomenclature for our simulations is straightforwardly obtained by combining either ``Q'' or ``P'' (indicative of the value of $w_x$) and a string which refers to the value of $\alpha_{xb}$.}
\label{tab:simulations}
\end{table}

\begin{table}
\centering
\begin{tabular}{c c}
\hline \hline \\[-1em]
Parameter & Value \\ [0.5ex] \hline\hline \\[-1em]
$H_0$ & $67.3\,{\rm km}/{\rm s}/{\rm Mpc}$ \\ \hline \\[-1em]
$\Omega_x$ & $0.722$ \\ \hline \\[-1em]
$\Omega_m$  & $0.278$ \\ \hline \\[-1em]
$\Omega_b$  & $0.044$ \\ \hline \\[-1em]
$\sigma_8 (z=0, \Lambda{\rm CDM})$ & $0.832$ \\
\hline
\hline
\end{tabular}
\caption{Values of the cosmological parameters used in our N-body simulations. All density parameters ($\Omega_x$, $\Omega_m$, and $\Omega_b$) are evaluated today. We use the best-fit values of the cosmological parameters as inferred from the \textit{Planck} legacy data release~\citep{Planck:2018vyg}.}
\label{tab:parameters}
\end{table}

\begin{figure}
\includegraphics[width=0.9\linewidth]{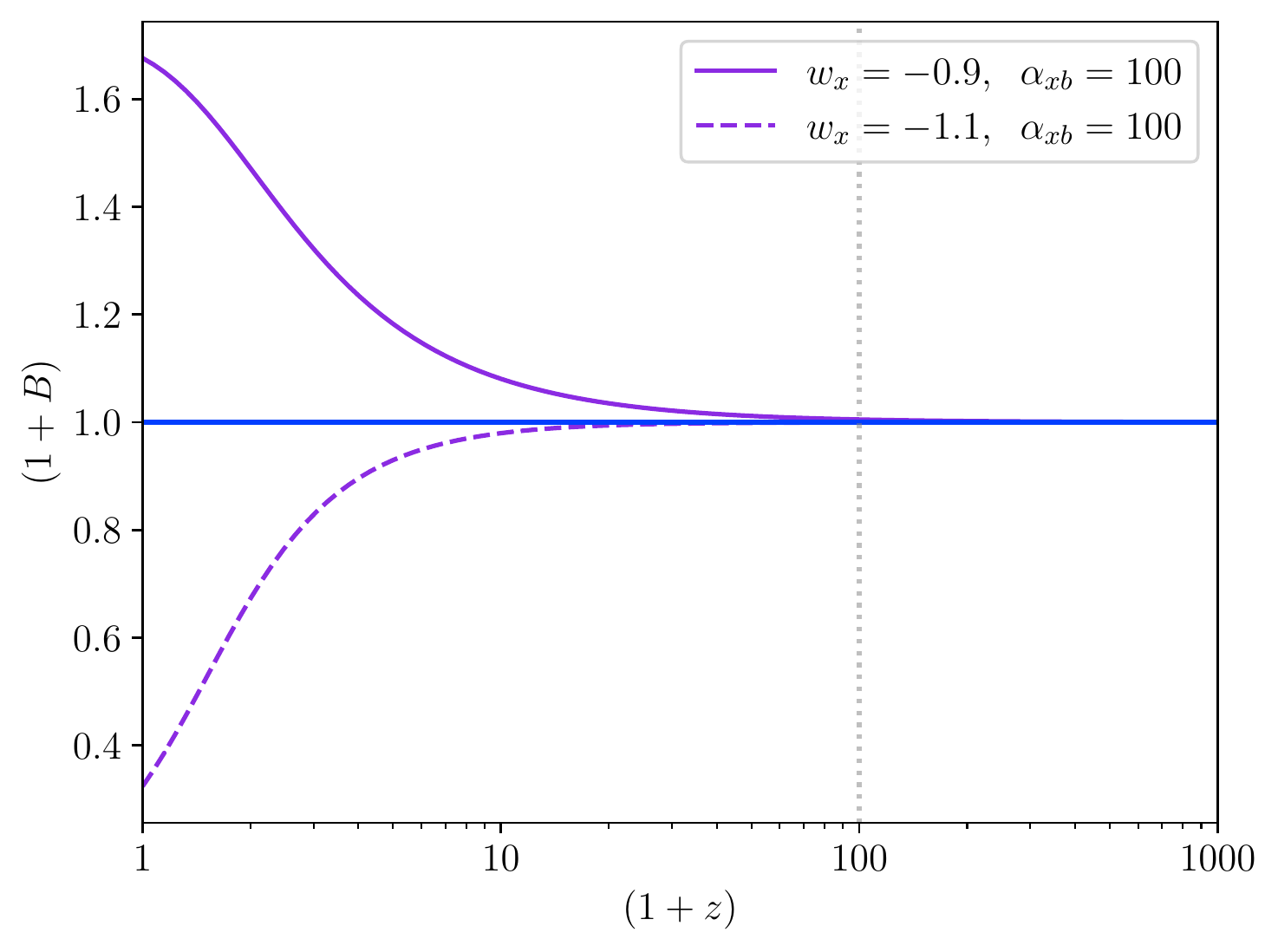}
\caption{Redshift evolution of the scattering factor $(1+B)$, with $B$ defined in Eq.~(\ref{eq:scattering}) and proportional to the dark energy-baryon scattering cross-section. We show the evolution of the scattering factor for the two most extreme cases we have simulated, with Thomson ratio given by $\alpha_{xb} = 100$, and dark energy equation of state given either by $w_x=-0.9$ (\textit{solid curve}, Q100 simulation) or $w_x=-1.1$ (\textit{dashed curve}, P100 simulation), in which case the effect of dark energy-baryon scattering leads to an extra friction or drag force respectively. We note that the effect of the scattering becomes rapidly negligible with increasing redshift, as dark energy becomes quickly subdominant. In particular at $z_i=99$ (denoted by the dotted vertical line), where we set the initial conditions for our N-body simulations, our assumption of $B \approx 0$ is valid to an extremely good approximation.}
\label{fig:scattering}
\end{figure}

To investigate signatures of momentum exchange between DE and baryons in the non-linear regime, we perform a series of multi-particle N-body simulations with a properly modified version of the tree-particle-mesh code \texttt{GADGET-3}~\citep{Springel:2000qu,Springel:2005mi}. We now discuss the implementation of the effects of DE-baryon scattering, as well as our simulation specifications.

Let us consider a generic system consisting of $N$ particles in an expanding universe. Taking the $i$th particle to be a DM particle with velocity $\boldsymbol{\varv}_i$, and considering the Newtonian limit, the gravitational acceleration it experiences is given by:
\begin{eqnarray}
\dot{\boldsymbol{\varv}}_i = -H{\boldsymbol{\varv}}_i + \sum_{j \neq i}^N \frac{Gm_j \mathbf{r}_{i,j}}{\vert \mathbf{r}_{i,j} \vert}\,,
\label{eq:dotvdm}
\end{eqnarray}
where $r_{i,j}$ is the separation between the $i$th and $j$th particles, $m_j$ is the mass of the $j$th particle, $H$ is the expansion rate, and $G$ is Newton's constant. On the other hand, if the $i$th particle is a baryonic one, it will experience an additional force due to scattering with DE, which depending on the sign of $(1+w_x)$ will act either as a friction ($+$ sign, quintessence-like DE) or a drag ($-$ sign, phantom DE). In this case, the gravitational acceleration is given by:~\footnote{Note that in a more realistic setting including screening mechanisms, as discussed at the end of Sec.~\ref{subsec:lagrangian}, this equation could be dynamically/environmentally modified based local quantities such as the SPH density.}
\begin{eqnarray}
\dot{\boldsymbol{\varv}}_i = - \left ( 1+B \right ) H{\boldsymbol{\varv}}_i + \sum_{j \neq i}^N \frac{Gm_j \mathbf{r}_{i,j}}{\vert \mathbf{r}_{i,j} \vert}\,,
\label{eq:dotvb}
\end{eqnarray}
where the $B$-dependent term encodes the scattering effects. We refer to $(1+B)$ as the ``scattering factor'', with $B$ being defined as:
\begin{eqnarray}
B \equiv (1+w_x) \alpha_{xb} \frac{c\sigma_{\textrm{T}}}{M_b}\frac{3\Omega_x}{8 \pi G}H\,.
\label{eq:scattering}
\end{eqnarray}
Here, $\Omega_x$ is the DE density parameter, and $M_b$ is the characteristic value of the baryonic particle mass, which in the present work we set equal to $0.5\,{\rm GeV}/c^2$, roughly corresponding to the average between the electron and proton masses $m_e$ and $m_p$ (i.e.\ $\approx m_p/2$, since $m_e \ll m_p$). We implement Eq.~(\ref{eq:dotvb}) in the acceleration equation for baryonic particles in \texttt{GADGET-3}, and have verified that our modified version of the code is essentially as fast as the standard one.

In writing Eq.~(\ref{eq:dotvb}), we have approximated the DE-baryon slip as $\Theta_{xb} \approx -\theta_b$, which is the reason why the DE velocity does not appear. In practice, this approximation is valid if the DE sound speed squared is $c_{s,x}^2 \approx 1$, which is the case for the simplest models of DE based on a single scalar field with a canonical kinetic term, minimally coupled to gravity, and in the absence of higher order operators (as in standard quintessence models). Under these assumptions, DE perturbations are damped on sub-horizon scales (i.e.\ those we are interested in for our N-body simulations) because the large pressure prevents the DE fluid from supporting perturbations efficiently, and the DE velocity field is to very good approximation homogeneous, so that we may safely approximate $\Theta_{xb} \approx -\theta_b$.

Of course, the assumption of a homogeneous DE component is just a zeroth-order approximation. In a more realistic setup, the scattering process would lead to the development of inhomogeneities in the DE and baryonic components, which should overall backreact on the background expansion rate. The size of this effect was estimated by one of us in the context of a related model in Sec.~2.3 of~\cite{Baldi:2014ica}, and was found to be negligible during the matter domination era, when the available kinetic energy is low, coupled to the fact that we take $w_x$ to be close to $-1$. We expect this estimate to carry on to our case as well~\citep[for further details see][]{Baldi:2014ica}. We plan to perform a more thorough investigation of this point in future follow-up work devoted to running more realistic simulations of DE-baryon scattering.

Each simulation in our suite features $512^3$ cold DM (CDM) particles of mass $7.57 \times 10^9\,h^{-1}M_{\odot}$ and $512^3$ collisionless baryonic particles of mass $1.42 \times 10^9\,h^{-1}M_{\odot}$. These are placed within a comoving box of size $250\,h^{-1}{\rm  Mpc}$. We save snapshots of our simulations from $z=2$ to $z=0$, separated by a redshift interval $\Delta z=0.5$. Moreover, we set the gravitational softening to $\epsilon_g = 12\,h^{-1}\,{\rm kpc}$ (to avoid large-angle scattering in two-body collisions), approximately corresponding to $1/40$th of the mean inter-particle separation.

It is worth noting that we do not include non-adiabatic processes such as radiative cooling, star and galaxy formation, feedback from Supernovae (SNe) and Active Galactic Nuclei (AGN), and so on, and that we basically treat baryons in the simulations as a separate family of collisionless particles. Therefore, our simulations can actually be re-interpreted as describing a model where only a fraction $f_{\rm DM} \approx 15\%$ of the DM component scatters with DE. We note that systems where a only a small fraction of the DM is subject to exotic dynamics (which include interactions with other components such as baryons, decays on cosmological timescales, and/or dissipative dynamics), have gained significant attention in the recent literature. We also note that our simulations differ significantly from the earlier DM-DE elastic scattering simulations of~\cite{Baldi:2014ica,Baldi:2016zom}, which only included the DM and DE fluids but not the baryonic one: therefore, these earlier simulations concerned a two-fluid cosmological system featuring scattering between DE and a collisionless fluid making up the rest of the energy budget. On the other hand, ours are simulations of a three-fluid cosmological system where DE and only one of the two remaining collisionless fluids (the subdominant one) are allowed to scatter.

For discussions on various approaches towards including non-adiabatic processes in hydrodynamical simulations, see e.g.~\cite{Sijacki:2007rw,vanDaalen:2011xb,Scannapieco:2011yd,Fabian:2012xr,Vogelsberger:2013eka,Vogelsberger:2014kha,Vogelsberger:2014dza,Mccarthy:2017yqf,Henden:2018som,Giri:2021qin}, as well as~\cite{Euclid:2020tff} and the cosmology-oriented review of~\cite{Chisari:2019tus}. The reason for not including these processes is two-fold: firstly, we do so in the interest of simplicity, in order to more cleanly isolate the signatures of DE-baryon scattering on the observables we consider, as this is the first ever investigation of the effects of DE-baryon scattering on the non-linear formation of cosmic structures. Next, and perhaps most importantly, particularly for large values of $\alpha_{xb}$ DE-baryon scattering can potentially have a significant impact on these non-adiabatic processes, to the extent that it is unclear whether we can simply make use of the standard ($\Lambda$CDM-driven) approaches towards including these processes adopted in standard simulations.

\begin{figure}
\includegraphics[width=1.0\linewidth]{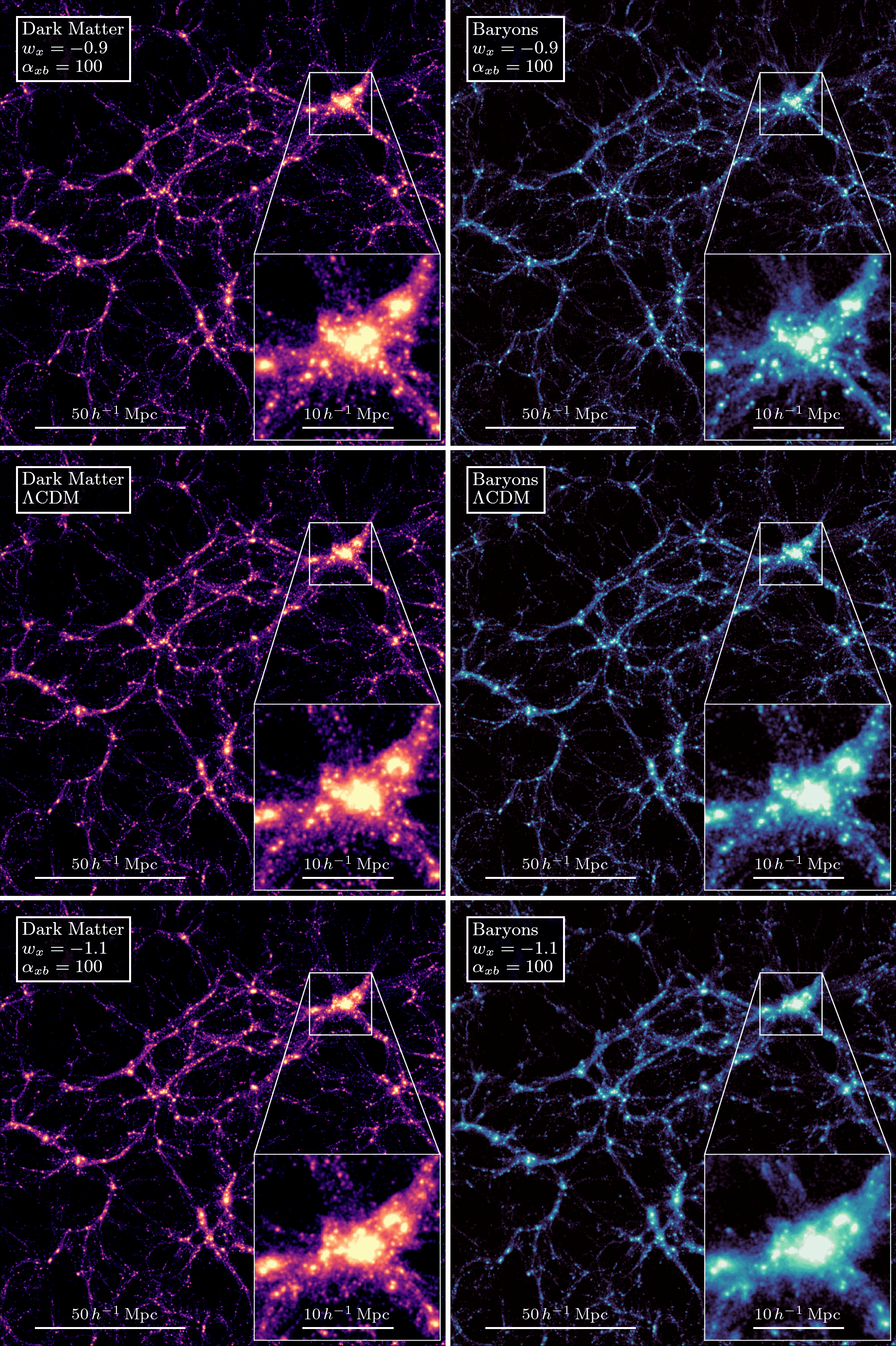}
\caption{Density slices from our simulations at $z=0$. The brighter the region, the larger the corresponding overdensity. \textit{Left column}: density slices from the cold dark matter density field, from the reference $\Lambda$CDM simulation where $w_x=-1$ and $\alpha_{xb}=0$ (middle panel), the Q100 simulation featuring dark energy-baryon scattering with Thomson ratio $\alpha_{xb}=100$ and dark energy equation of state $w_x=-0.9$ (top panel), and the P100 simulation with $\alpha_{xb}=100$ and $w_x=-1.1$ (bottom panel). \textit{Right column}: as in the left column, but focusing on the baryon density field.}
\label{fig:density}
\end{figure}

We run a suite of 9 simulations, characterized by the chosen values of the DE EoS $w_x$ and the Thomson ratio $\alpha_{xb}$. Of these 9 simulations, one is a reference $\Lambda$CDM simulation, i.e.\ where $w_x=-1$ and $\alpha_{xb}=0$. We use this as baseline against which we compare the results of all our other simulations. We then fix the DE EoS $w_x$ to two values: in one case $w_x=-0.9$ within the quintessence-like regime (``Q''), and in the other case $w_x=-1.1$ within the phantom regime (``P'').~\footnote{These two values are selected as being representative of the maximum value of $\vert 1+w_x \vert \sim 0.1$ allowed by current observational constraints on $w_x$~\citep{eBOSS:2020yzd}, thus maximizing the size of the effects associated with DE-baryon scattering, which are parametrically controlled by $(1+w_x)$.} Concerning the Thomson ratio $\alpha_{xb}$, we fix it to four values: $\alpha_{xb}=0\,,1\,,10\,,100$. The $\alpha_{xb}=0$ case corresponds to a reference no-scattering $w$CDM model (with $w_x \neq -1$), that we also use as a baseline against which we compare the results of all the simulations with $\alpha_{xb} \neq 0$ (at fixed $w_x$). On the other hand, values of $\alpha_{xb} \sim 1-10$ were shown to leave no visible effect on linear cosmological observables in~\cite{Vagnozzi:2019kvw}, but could potentially be phenomenologically justified by the latest XENON1T results~\citep{XENON:2020rca}, if interpreted in terms of a direct detection of chameleon-screened DE, as shown in~\cite{Vagnozzi:2021quy}. Aside from the reference $\Lambda$CDM simulation, our other 8 simulations are obtained by combining the 2 selected values of $w_x$ and 4 selected values of $\alpha_{xb}$. The nomenclature for our simulations is summarized in Table~\ref{tab:simulations}, and is straightforwardly obtained by combining either ``Q'' or ``P'' (indicative of the value of $w_x$) and a string which refers to the value of $\alpha_{xb}$: for instance, Q1 denotes a simulation with $w_x=-0.9$ and $\alpha_{xb}=1$, and similarly P100 indicates a simulation with $w_x=-1.1$ and $\alpha_{xb}=100$.

As discussed earlier, our phenomenological DE-baryon elastic scattering model only affects the evolution of perturbations through momentum exchange, but not the background evolution, since there is no energy exchange. Because of this, it is possible to use the same initial conditions for each simulation despite the fact that these have different values of the Thomson ratio $\alpha_{xb}$. It should be noted that this choice results in different values of $\sigma_8$ at $z=0$ depending on the value of $\alpha_{xb}$. We generate initial conditions by means of the \texttt{N-GenIC} code~\citep{Springel:2005mi}. This is based on the Zeldovich approximation~\citep{Zeldovich:1969sb}, and displaces particles from a homogeneous Cartesian lattice in such a way that their density distribution will correspond to a specific random-phase realisation of the matter power spectrum. We compute the matter power spectrum employed to set the initial conditions through the publicly available Boltzmann solver \texttt{CAMB}~\citep{Lewis:1999bs}, assuming a $\Lambda$CDM cosmology with cosmological parameters given in Table~\ref{tab:parameters}. This choice is consistent with the best-fit values of the cosmological parameters inferred from the \textit{Planck} CMB legacy data release~\citep{Planck:2018vyg}.

We choose $z_i=99$ as starting redshift, excluding any possibility of scattering between DE and baryons for $z>z_i$. In our case, this assumption is extremely reasonable. In fact, the extra scattering-dependent term in Eq.~(\ref{eq:dotvb}) depends on the DE density parameter $\Omega_x$, which becomes quickly negligible with increasing redshift, when DE is a subdominant component of the Universe. We show this explicitly in Fig.~\ref{fig:scattering}, where we plot the evolution with redshift of the scattering factor $(1+B)$ [see Eqs.~(\ref{eq:dotvb},\ref{eq:scattering})] for the two most extreme cases: $\alpha_{xb} = 100$, with $w_x$ fixed to either $w_x=-0.9$ (solid curve) or $w_x=-1.1$ (dashed curve). We see that already at $z \sim 10$, $B \ll 1$, and the scattering factor is essentially indistinguishable from unity at $z=z_i$, which is where our initial conditions are set. The fact that DE-baryon scattering is completely negligible at high redshift also justifies our choice of setting initial conditions using a $\Lambda$CDM power spectrum, as the latter would in any case be totally indistinguishable from the power spectrum within the DE-baryon scattering model, at least for the region of parameter space we have chosen to explore.

\section{Results}
\label{sec:results}

Here we discuss the main results of our work, focusing on relevant cosmological observables which we extract from our N-body simulations discussed in Sec.~\ref{sec:simulations}. After general considerations on the large-scale density distribution, we discuss the non-linear matter power spectrum, halo mass function, halo profiles, and halo baryon fraction profiles. In particular, to isolate the impact of DE-baryon scattering on such observables, we will compare simulations including the effect of scattering versus $w$CDM simulations with no scattering but $w_x \neq -1$ (i.e. the Q0 and P0 simulations), as well as versus the reference $\Lambda$CDM simulation with $w_x=-1$ and $\alpha_{xb}=0$.

\subsection{Large-scale density distribution}
\label{subsec:distribution}

We start our analysis by providing general qualitative considerations on the large-scale density distribution observed in our N-body simulations. To do so, we consider representative density slices of our simulations, computed through a Cloud-in-Cell (CIC) scheme. In Fig.~\ref{fig:density} we show density slices of the baryons and CDM fields at $z=0$ for three different simulations: the reference $\Lambda$CDM simulation, and the two simulations including the effects of the most extreme scattering process ($\alpha_{xb}=100$), namely the Q100 and P100 simulations. Each density slice covers a box of volume $(150 \times 150 \times 15)\,h^{-3}{\rm Mpc}^3$. The overall shape of the cosmic web on the largest scales is visibly the same for every simulation, reflecting the fact that they all share the same initial conditions.

Comparing the CDM density distributions, no significant differences can be noticed, at least by the naked eye, in agreement with expectations. On the other hand, comparing the density distributions of baryons, we observe that the Q100 simulation with $w_x=-0.9$ and $\alpha_{xb}=100$ displays an overall suppressed clustering of the highest density peaks with respect to the $\Lambda$CDM case. Conversely, we notice the opposite trend for the P100 simulation with $w_x=-1.1$ and $\alpha_{xb}=100$, wherein the highest density peaks are more strongly clustered. This effect can be seen more clearly in the zoomed panels of each plot, centered on the highest density peak of the specific simulation slice. This behaviour matches the results reported in~\cite{Vagnozzi:2019kvw}, in agreement with our expectations on the effects of DE-baryon scattering on the linear matter power spectrum: suppressed or enhanced power depending on whether $w_x>-1$ or $w_x<-1$ respectively, as the former leads to a friction force slowing down the growth of structure on the largest scales, whereas the latter leads to a drag force which enhances the growth of structure on the same scales.

In summary, a qualitative assessment of the large-scale density distribution extracted from our simulations confirms the linear results already presented in~\cite{Vagnozzi:2019kvw}, as well as our understanding of the linear effects of DE-baryon scattering in terms of an extra friction/drag force for scattering with a quintessence-like/phantom DE component. Having confirmed that linear structure formation proceeds as expected, we can now turn our attention to the non-linear formation of cosmic structures.

\subsection{Non-linear power spectra}
\label{subsec:powerspectrum}

\begin{figure}
\includegraphics[width=0.9\linewidth]{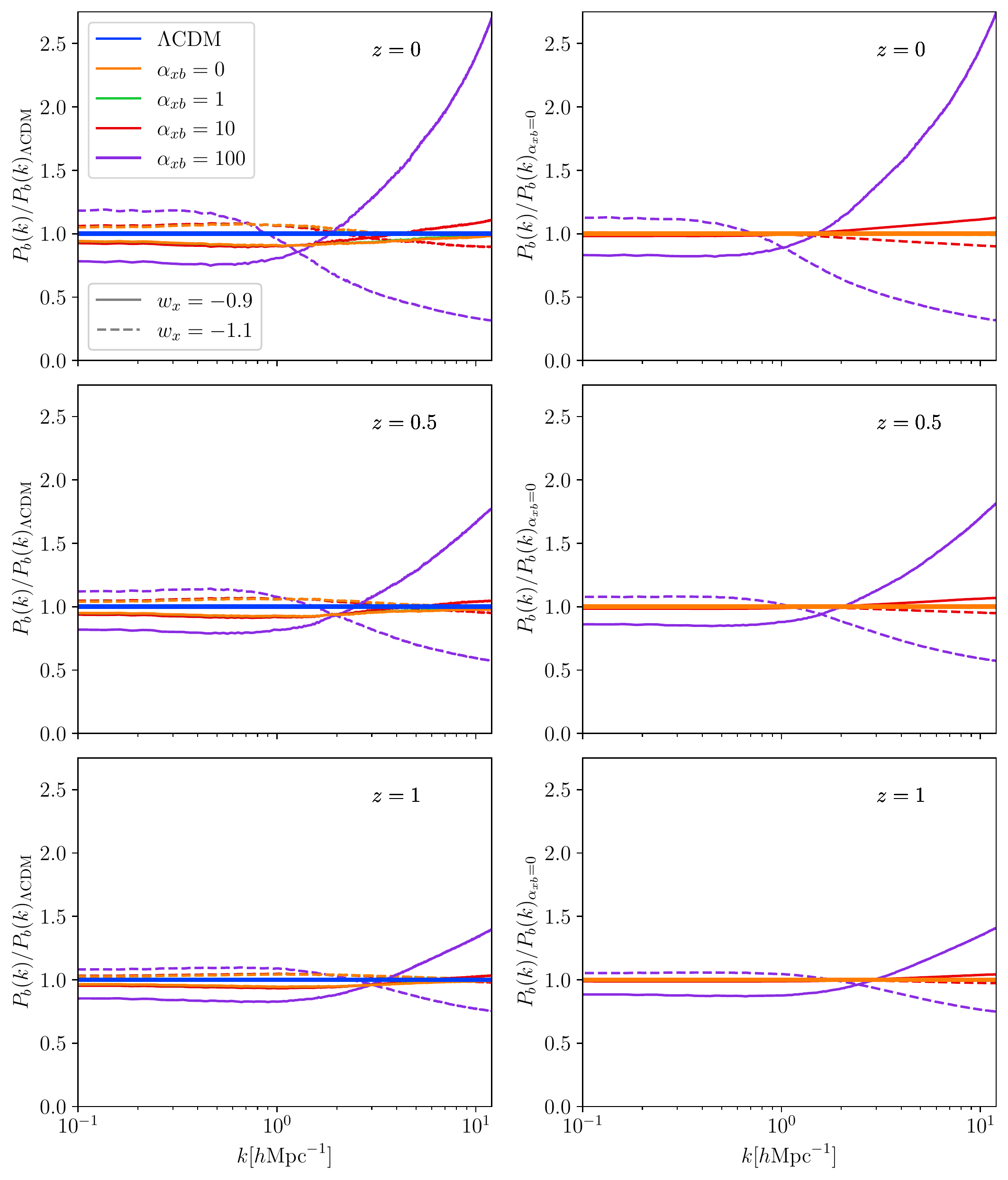}
\caption{\textit{Left column}: ratio of the baryon power spectrum in models featuring dark energy-baryon scattering relative to the reference $\Lambda$CDM power spectrum where $w_x=-1$ and $\alpha_{xb}=0$, at $z=0$ (top panel), $z=0.5$ (middle panel), and $z=1$ (bottom panel). Within each panel, we consider various values of the Thomson ratio $\alpha_{xb}$ as determined by the color coding, whereas the dark energy equation of state is given either by $w_x=-0.9$ (solid curves) or $w_x=-1.1$ (dashed curves). \textit{Right column}: as in the left column, with the same color coding for the different values of $\alpha_{xb}$, but this time considering the ratio relative to the reference no-scattering $w$CDM power spectra where $\alpha_{xb}=0$ and the dark energy equation of state is given either by $w_x=-0.9$ (solid curves) or $w_x=-1.1$ (dashed curves). We clearly see that the direction of the deviations relative to the reference models depend on the sign of $(1+w_x)$, and that the deviations on non-linear scales go in the opposite direction relative to their linear counterparts.}
\label{fig:pk_ratio_baryons}
\end{figure}

\begin{figure}
\includegraphics[width=0.9\linewidth]{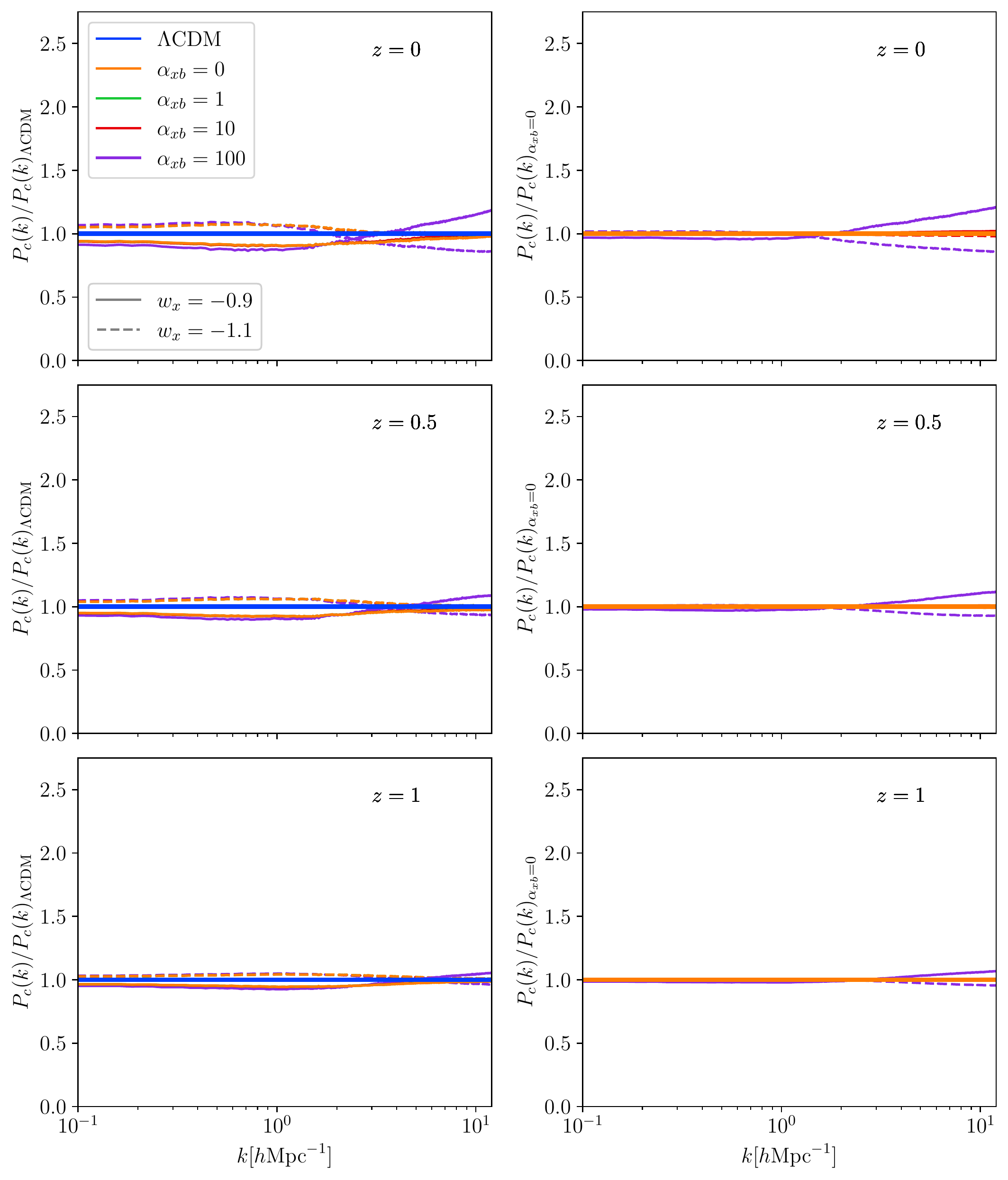}
\caption{As in Fig.~\ref{fig:pk_ratio_baryons}, but focusing on the cold dark matter power spectrum.}
\label{fig:pk_ratio_cdm}
\end{figure}

\begin{figure}
\includegraphics[width=0.9\linewidth]{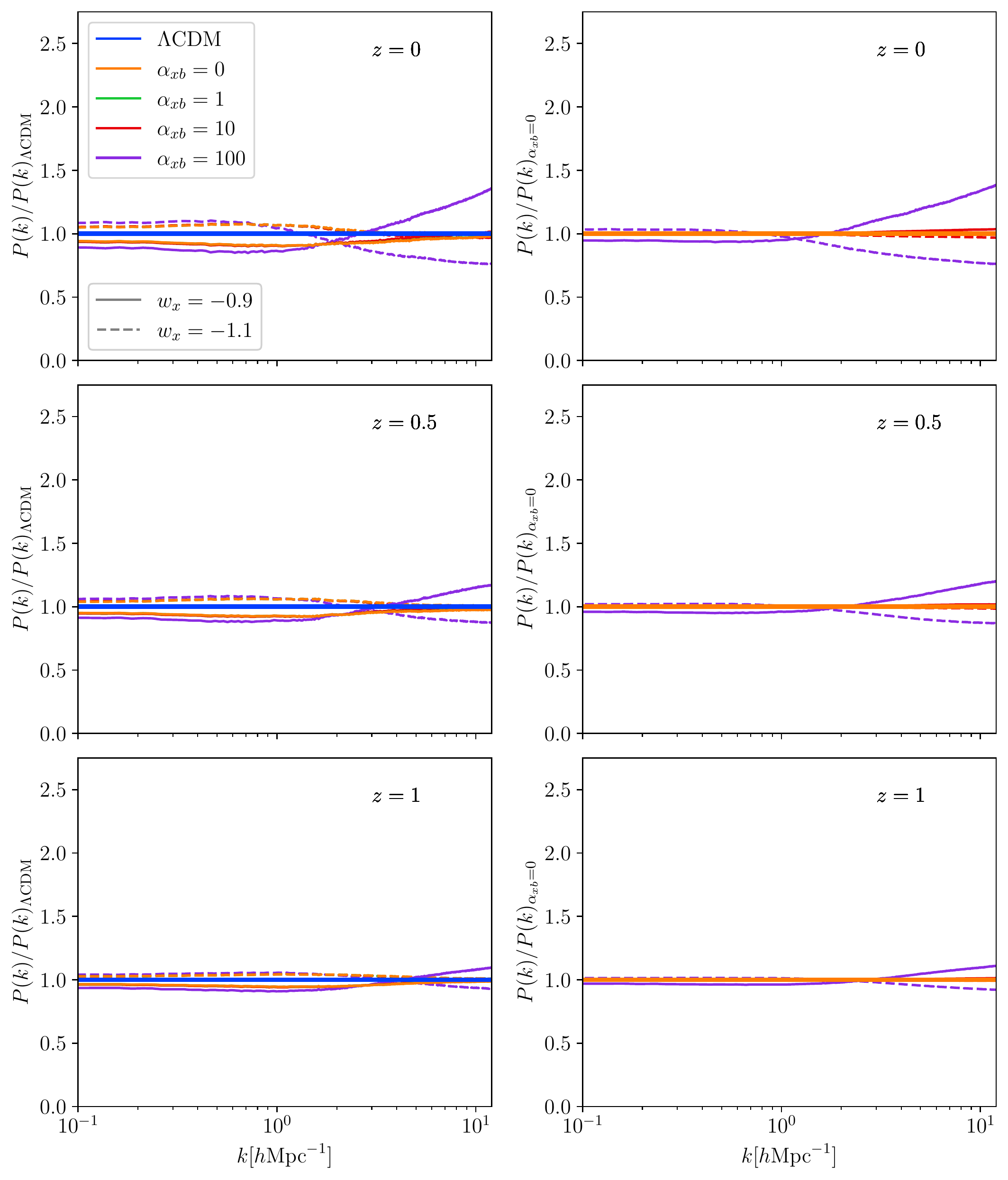}
\caption{As in Fig.~\ref{fig:pk_ratio_baryons}, but focusing on the power spectrum of the total matter field (baryons and cold dark matter).}
\label{fig:pk_ratio_matter}
\end{figure}

We now compute the power spectra of the baryons, CDM, and total matter fields, at three reference redshifts: $z=0\,,0.5\,,1$. The spectra have been computed by assigning the mass to a cubic Cartesian grid by means of the CIC scheme: the chosen grid has half the spacing of the mesh used for the large-scale N-body integration, corresponding to $1024^3$ modes. We can reliably extract the power spectrum up to a wavenumber corresponding to the Nyquist frequency associated with the grid, which is given by $k_{\rm Ny} = \pi N/L \approx 12.9\,h{\rm Mpc}^{-1}$. We stress that what we are extracting here are the \textit{non-linear} power spectra, measured directly from our N-body simulations.

We plot the ratios between the power spectra obtained from each of these simulations relative to the reference $\Lambda$CDM power spectra, in the left columns of Figs.~\ref{fig:pk_ratio_baryons} (baryons), \ref{fig:pk_ratio_cdm} (CDM), and \ref{fig:pk_ratio_matter} (total matter). However, to more fairly isolate the direct perturbation effects of DE-baryon scattering from effects caused by a change in the background expansion, the right columns of the same Figures compare these power spectra to the power spectra of a reference no-scattering $w$CDM model, i.e.\ one with the same value of the DE EoS $w_x \neq -1$, but with $\alpha_{xb}=0$. For example, the Q1, Q10, and Q100 power spectra are all compared to the Q0 power spectra, and similarly for the P1, P10, and P100 cases which are compared to the P0 case. Here and in all subsequent plots, solid curves refer to scattering with a quintessence-like DE component, while dashed curves are associated with    a phantom DE component, and the colors of the curves determine the value of $\alpha_{xb}$.

First of all, we notice from Figs.~\ref{fig:pk_ratio_baryons}--\ref{fig:pk_ratio_matter} that the effects of DE-baryon scattering become more significant as we approach $z=0$. One reason is that the effects of the scattering at low redshift have had a longer time over which to accumulate. Moreover, from Fig.~\ref{fig:scattering}, we see that the scattering factor $(1+B)$ increases significantly with time, reflecting the fact that DE becomes dynamically more important, and eventually dominant, as we approach late times. From Figs.~\ref{fig:pk_ratio_baryons}--\ref{fig:pk_ratio_matter} we also see that the effects of DE-baryon scattering are reversed in sign when considering $w_x=-0.9$ versus $w_x=-1.1$. This reflects the fact, amply discussed previously and reflected in both the Boltzmann equations [Eq.~(\ref{eq:thetab})] and the extra force interpretation [Eq.~(\ref{eq:f})], that the direction in which the effects of DE-baryon scattering operate depends on the sign of $(1+w_x)$, or equivalently the sign of the quantity $B$ [Eq.~(\ref{eq:scattering})].

Let us focus on Fig.~\ref{fig:pk_ratio_baryons}, which shows the effect of DE-baryon scattering on the non-linear power spectrum of baryons. The linear effects consist in an suppression or enhancement of power for $w_x>-1$ or $w_x<-1$ respectively, in agreement with the physical interpretation provided earlier in~\cite{Vagnozzi:2019kvw}. In the linear regime, the peculiar velocities of baryonic particles are aligned with the gradient of the potential, and therefore DE-baryon scattering acts as a friction [drag] term suppressing [enhancing] structure formation in the $w_x>-1$ [$w_x<-1$] cases respectively. On linear scales the effect leads to a roughly scale-independent change in the power spectrum which is well approximated by $\vert \Delta P(k)/P(k) \vert \approx 1.5 \times 10^{-3}\alpha_{xb}$ at $z=0$, and is smaller at higher $z$. For the most extreme Q100 and P100 simulations (with $\alpha_{xb}=100$), the deviation from the no-scattering $w$CDM model is in fact $\approx 15\%$ at $z=0$, slightly decreasing as $z$ increases in the Q100 case, with the redshift evolution being somewhat more significant in the P100 case. The changes in large-scale power can be interpreted as shifts in $\sigma_8$, which for $\alpha_{xb} \sim 100$ and $w_x>-1$ are large enough [${\cal O}(10\%)$] and go in the right direction to potentially be of interest in the context of the $S_8$ discrepancy~\citep[see e.g.][]{DiValentino:2018gcu,DiValentino:2020vvd,Nunes:2021ipq}: we defer a full investigation of this issue to future work.

More interesting results are found when we focus on non-linear modes. In this case, we observe a strong transition in the behavior of the matter power spectrum, typically occurring within the wavenumber range $0.6\,h{\rm Mpc}^{-1} \lesssim k \lesssim 2\,h{\rm Mpc}^{-1}$. Beyond the transition, the relative changes in the power spectrum with respect to both the $\Lambda$CDM and no-scattering $w$CDM models not only switch sign, but increase significantly in amplitude. In other words, we observe an enhancement [suppression] of power in the $w_x>-1$ [$w_x<-1$] cases respectively, i.e.\ a reversed trend compared to that observed in the linear regime.

To interpret this behavior, it is useful to borrow the earlier results obtained by one of us in~\cite{Baldi:2014ica,Baldi:2016zom} in relation to the DM-DE elastic scattering model. In fact, as already mentioned above, in the linear regime the baryon peculiar velocity field is always aligned with the spatial gradient of the gravitational potential, implying that the extra force due to DE-baryon scattering always acts in the same direction of the gravitational acceleration. However, this is no longer true in the non-linear regime, i.e.\ after shell crossing, due to the fact that collapsing structures gain angular momentum. In this regime, an extra friction [drag] force supports [opposes] the loss of angular momentum by baryonic particles in bound structures. This significantly alters the virial equilibrium of bound structures, causing them to contract [expand], which in turn results in a faster [slower] collapse of non-linear structure, consequently enhancing [suppressing] the efficiency of non-linear structure formation and consequently the power spectrum. To put it differently, in the non-linear regime an extra friction [drag] force dissipates [injects] kinetic energy from [into] the system, suppressing [enhancing] structure formation, which is the exact opposite of what happens in the linear regime.

Therefore, while scattering between baryons and a quintessence-like [phantom] DE component suppresses [enhances] power on large scales, the reverse occurs on small scales, due to the important role of angular momentum in collapsing structures. We notice that non-linear effects become significant also for the previously undetectable $\alpha_{xb} = 10$ case. Moreover, the redshift dependence of these effects is significantly stronger than their linear counterparts. At a wavenumber $k \sim 10\,h{\rm Mpc}^{-1}$, the relative changes with respect to the no-scattering $w$CDM model are $\approx 140\%$ at $z=0$, $\approx 70\%$ at $z=0.5$, and $\approx 35\%$ at $z=1$ for the most extreme case with $\alpha_{xb}=100$, with the deviations being approximately one order of magnitude smaller for the less extreme case with $\alpha_{xb}=10$. On the other hand, even within the non-linear regime the $\alpha_{xb}=1$ case DE-baryon scattering does not appear to lead to visible signatures with respect to the no-scattering $w$CDM model, and remains challenging to probe.

In Fig.~\ref{fig:pk_ratio_cdm} we instead consider the CDM power spectra. Compared to the baryon power spectra, all the effects previously identified still appear, albeit significantly suppressed. This is to be expected, as CDM is not scattering with DE in our model. Therefore, the effects of DE scattering on the power spectra are only indirectly transmitted to the CDM component via gravitational interactions with baryons, the latter however being a subdominant component. Within the linear regime, the relative deviations in the CDM power spectra compared to the no-scattering reference $w$CDM model never exceed $\approx 3\%$, even for the most extreme cases ($\alpha_{xb}=100$). On the other hand, in the non-linear regime we observe significant deviations when $\alpha_{xb}=100$: at wavenumbers $k\sim 10\,h{\rm Mpc}^{-1}$, the relative deviations are $\approx 20\%$ at $z=0$, $\approx 10\%$ at $z=0.5$, and $\approx 6\%$ at $z=1$. While not reaching the $\gtrsim 100\%$ relative changes observed earlier in the baryon power spectrum, these changes are still large enough to be potentially visible, albeit with caveats to be discussed later. As with the baryon power spectra, the physical explanation for these deviations in the CDM power spectra can again be traced to the role of angular momentum in collapsing structures, with DE-baryon scattering altering the virial equilibrium thereof (and the corresponding deviations in the CDM component being transmitted from the baryon component via gravitational interactions between baryons and CDM).

Finally, in Fig.~\ref{fig:pk_ratio_matter} we show the relative deviations in the power spectra of the total matter (baryons+CDM) component. Matter fluctuations are given by a weighted average of the baryon and CDM fluctuations [$\delta_m \approx (\Omega_c\delta_c + \Omega_b\delta_b)/(\Omega_c+\Omega_b)$]. Therefore, we expect the size of the effects of DE-baryon scattering on the total matter power spectra to be intermediate between the sizes of the effects previously seen on the baryons and CDM power spectra. Our simulations do indeed confirm this expectation. We see that at the linear level, when $\alpha_{xb} = 100$, there is a relative enhancement/suppression with respect to the no-scattering reference $w$CDM model which is as large as $\approx 10\%$ at $z=0$, and $\approx 4\%$ at $z=1$. As before, the non-linear effects are more significant: at scales $k\sim 10\,h{\rm Mpc}^{-1}$ and for $\alpha_{xb} = 100$, the relative deviation is this time $\approx 30\%$ at $z=0$, $\approx 15\%$ at $z=0.5$, and $\approx 10\%$ at $z=1$. For $\alpha_{xb}=10$ the deviations are significantly more modest: for instance, at scales $k\sim 10\,h{\rm Mpc}^{-1}$ and at $z=0$, the relative deviation is $\approx 3\%$, which is challenging but not impossible to probe.

These relative deviations are larger than those observed in the CDM power spectra, although nowhere close to the $\gtrsim 100\%$ deviations observed in the baryon power spectra. The reason has to do with the fact that matter fluctuations are a weighted average of baryons and CDM fluctuations. At the level of power spectra (i.e.\ two-point functions of the density fluctuations), the CDM component is up-weighed by a factor of $\approx (\Omega_c/\Omega_b)^2 \approx 25$, which explains the size of the effects observed. However, we note that for the purpose of direct comparison to observations, it is the total matter power spectrum which is of direct relevance, not the power spectra of the individual baryons or CDM components. In fact, it is the total matter power spectrum which is accessible to weak lensing measurements or to LSS clustering measurements (in the latter case up to factors associated to LSS tracer bias).

In summary, we have found that DE-baryon scattering leaves an important imprint in the non-linear clustering of the LSS. In particular, the deviations with respect to the reference models on non-linear scales go in the opposite direction compared to their linear counterparts, and are significantly larger than the latter. This suggests that probes of non-linear structure formation may be able to set significantly tighter constraints on the DE-baryon scattering cross-section compared to linear observables, provided of course one can reliably model non-linear theoretical predictions to the precision required by next-generation surveys.

\subsection{Halo mass function}
\label{subsec:halomassfunction}

\begin{figure}
\includegraphics[width=0.9\linewidth]{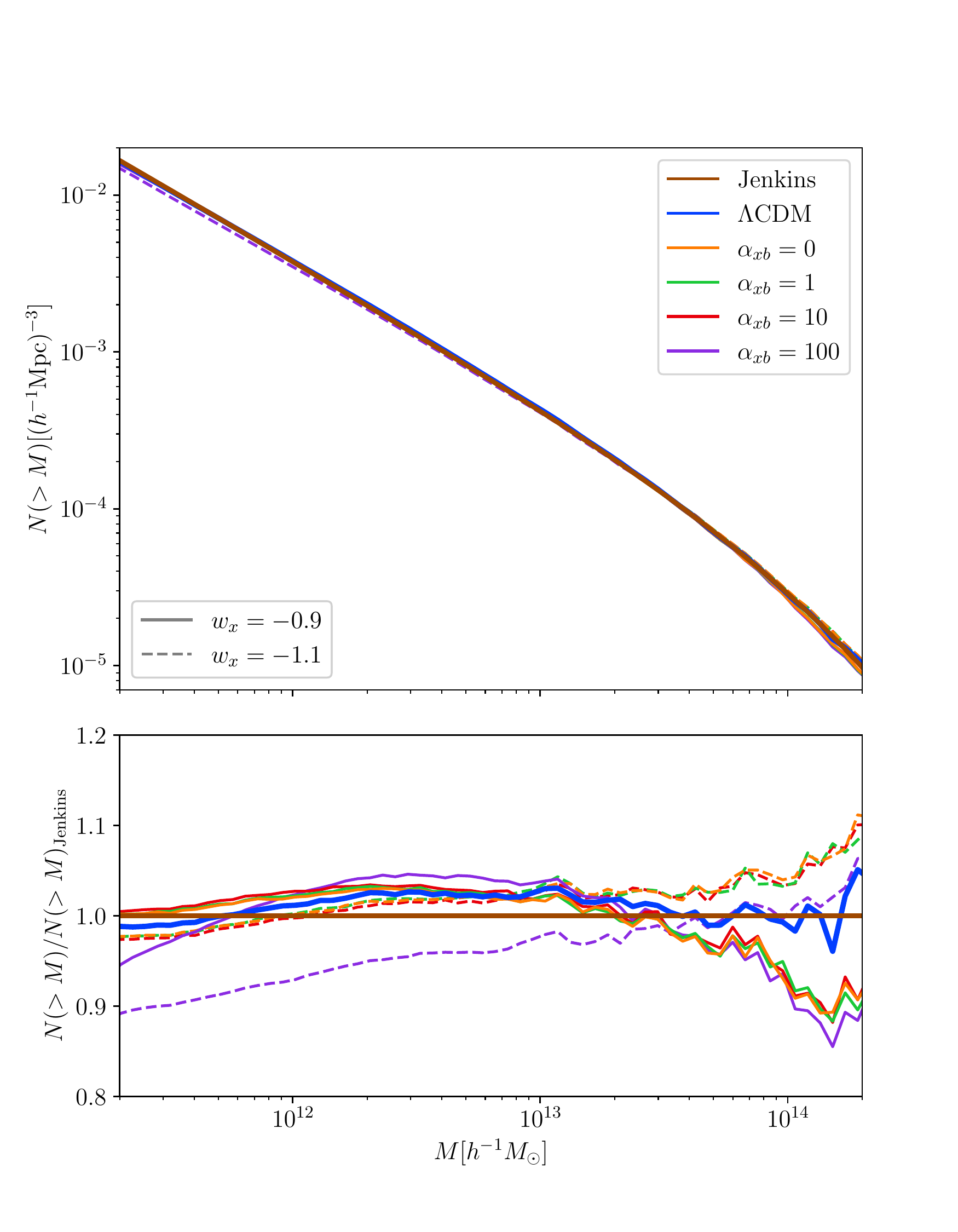}
\caption{\textit{Upper panel}: cumulative halo mass function at $z=0$ as predicted by Jenkins~\citep[brown curve,][]{Jenkins:2000bv}, and extracted from our simulations both within the reference $\Lambda$CDM model (blue curve) as well as within models featuring dark energy-baryon scattering with Thomson ratio $\alpha_{xb}$ determined by the color coding, whereas the dark energy equation of state is given by $w_x=-0.9$ (solid curves) or $w_x=-1.1$ (dashed curves). \textit{Lower panel}: ratios of the various cumulative mass functions plotted in the upper panel relative to the prediction from the Jenkins mass function. We also note that our reference $\Lambda$CDM model (solid blue curve) is in excellent agreement with the prediction from the Jenkins mass function: while the ratio between the two exhibits a slight scale dependence, the relative deviations always remain $\lesssim 3\%$.}
\label{fig:mass_function}
\end{figure}

\begin{figure}
\includegraphics[width=0.9\linewidth]{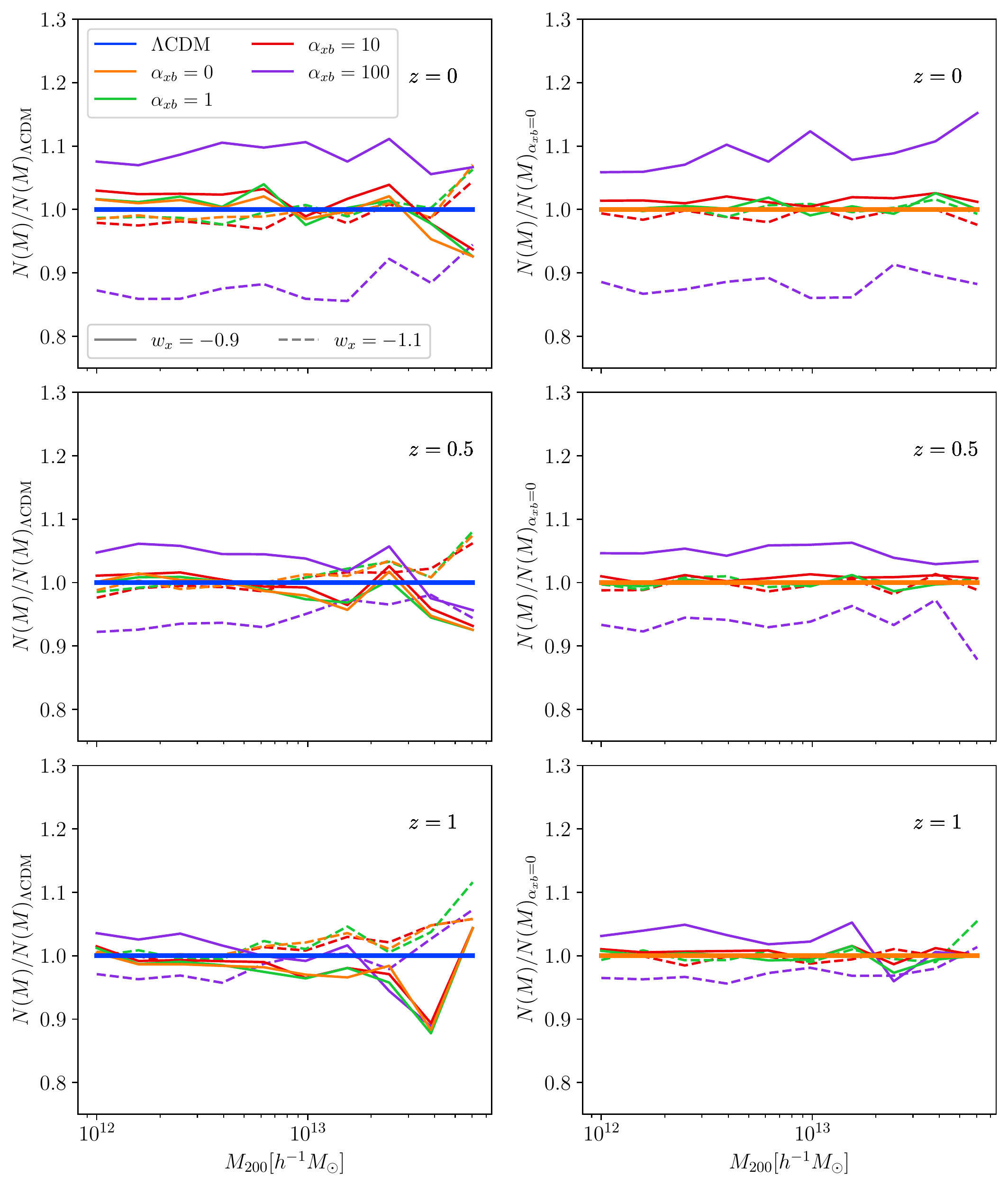}
\caption{As in Fig.~\ref{fig:pk_ratio_baryons} but focusing on the halo mass function. Note that this is not the cumulative halo mass function considered in Fig.~\ref{fig:mass_function}.}
\label{fig:mass_function_ratio}
\end{figure}

As an additional observable beyond the power spectra of the baryons, CDM, and matter components, we consider the halo mass function $N(M)$, i.e.\ the number density of halos of a given mass:
\begin{eqnarray}
N(M) \equiv dN/dMdV\,.
\end{eqnarray}
More specifically, we compute the halo mass function as the number of halos of virial mass $M_{200}$ lying within a set of logarithmically spaced mass bins, where $M_{200}$ is defined as the mass enclosed within a sphere of virial radius $R_{200}$ constructed around the center of a halo, such that the halo density is $200$ times the critical density of the Universe $\rho_{\rm crit} \equiv 3H_0^2/8\pi G$:
\begin{eqnarray}
\frac{M_{200}}{4\pi R_{200}^3} = 200\rho_{\rm crit}\,.
\label{eq:m200}
\end{eqnarray}
We also note that the choice of setting the threshold to $200\rho_{\rm crit}$ is simply a matter of convenience, for the ease of comparison to related works~\citep[see e.g.][]{Navarro:1995iw}.

We generate halo catalogues from our N-body simulations using the \texttt{SUBFIND} routine~\citep{Springel:2000qu}, which identifies halos as overdense, self-bound groups of particles within a larger parent group by means of gravitational unbinding. As a first consistency check, we consider the cumulative mass function $N(>M)$, which is related to the probability of finding a given halo with mass greater than $M$:
\begin{eqnarray}
N(>M) = \int_M^{\infty} d\tilde{M}\,N(\tilde{M})\,.
\label{eq:cumulative}
\end{eqnarray}
In Fig.~\ref{fig:mass_function} we compare the cumulative mass function within our models featuring DE-baryon scattering, against the cumulative Jenkins mass function predicted by~\cite{Jenkins:2000bv} within the $\Lambda$CDM model (with the same choice of cosmological parameters). While the ratio between the two cumulative mass functions exhibits a slight scale-dependence, the relative deviation from the Jenkins mass function is never large: for instance, within the halo mass range $2 \times 10^{11} \lesssim M/(h^{-1}M_{\odot}) \lesssim 10^{14}$, the relative deviation never exceeds $\lesssim 10\%$ even in the most extreme case where $\alpha_{xb}=100$, and is significantly smaller ($\lesssim 3\%$) for $\alpha_{xb}=10$.

In Fig.~\ref{fig:mass_function_ratio} we plot the ratios between the (non-cumulative) halo mass functions within each of our simulations relative to the reference $\Lambda$CDM model in the left column, and relative to the no-scattering reference $w$CDM model in the right column. These ratios are computed at redshifts $z=0\,,0.5\,,1$. To produce these plots, we have used 10 logarithmically equispaced mass bins in the halo mass range $2 \times 10^{11} \lesssim M/(h^{-1}M_{\odot}) \lesssim 10^{14}$.

As before, we notice that the effects of DE-baryon scattering become significantly more prominent the more we approach $z=0$. For the most extreme cases, i.e. the Q100 and P100 simulations with $\alpha_{xb} = 100$, we find non-negligible deviations from both the reference $\Lambda$CDM and $w$CDM models, with relative deviations as large as $\approx 15\%$. For the Q100 case ($w_x=-0.9$) at $z=0$, we observe a $\approx 10-15\%$ mass-independent enhancement in the abundance of halos at all masses. Similarly, for the P100 case ($w_x=-1.1$) at $z=0$, we find a $\approx 10-15\%$ mass-independent suppression of the abundance of halos at all masses. For the remaining cases with $\alpha_{xb}=10$ and $\alpha_{xb}=1$, we find much more modest relative deviations at the $\lesssim 1-3\%$ level at best, with the largest deviations being found for the $\alpha_{xb}=10$ case at $z=0$.

The physical explanation for these results is directly related to the explanation of the results observed when considering the non-linear power spectra in Sec.~\ref{subsec:powerspectrum}. In fact, halo masses are tied to non-linear collapse, and therefore respond to the effects observed in the non-linear power spectra. As we previously discussed, the extra friction force felt by the baryonic component will support the loss of angular momentum by bound structures, thereby easing non-linear halo collapse: this explains why the halo mass function is enhanced by scattering between baryons and a quintessence-like DE component. The converse is true for a phantom DE component, whose associated drag force opposes the loss of angular momentum by bound structures, thereby hindering non-linear halo collapse. These arguments hold for sufficiently low halo masses. However, the high mass tail of the halo mass function is exponentially sensitive to the amplitude of linear perturbations, and therefore responds also to the effects observed on the linear power spectra~\citep{Press:1973iz}. The amplitude of perturbations on linear scales is captured by $\sigma_8$, which we recall decreases [increases] in the presence of scattering with a quintessence-like [phantom] DE component. This explains the slight change of trend observed in the high mass ends of Figs.~\ref{fig:mass_function} and~\ref{fig:mass_function_ratio} (particularly the former, as the cumulative mass function carries integrated sensitivity to these effects).

In summary, DE-baryon scattering leaves a significant imprint in the halo mass function, which are a direct consequence of the signatures observed in the clustering of the LSS discussed previously. The signatures we have found are quite distinctive as their mass dependence is relatively weak. This can be contrasted to several beyond-$\Lambda$CDM ingredients, which typically impact the halo mass function (and structural properties of halos which we shall discuss later) in a strongly mass-dependent way, such as modified gravity~\citep{Lombriser:2013wta,Baldi:2013iza}, primordial non-Gaussianity~\citep{Desjacques:2009jb,LoVerde:2011iz,Yokoyama:2011sy}, self-interacting DM~\citep{Cyr-Racine:2013fsa,Foot:2014uba,Schneider:2014rda,Foot:2016wvj,Murgia:2017lwo,Lovell:2017eec,Bohr:2021bdm}, warm DM~\citep{Schneider:2013ria,Angulo:2013sza,Parimbelli:2021mtp}, and massive neutrinos~\citep{Costanzi:2013bha,Vagnozzi:2017ovm,Hagstotz:2018onp,Zennaro:2019aoi,Cataneo:2019fjp}, typically in response to strong scale-dependent modifications to the non-linear power spectrum.

\begin{figure*}
\begin{center}
\includegraphics[width=2.8in]{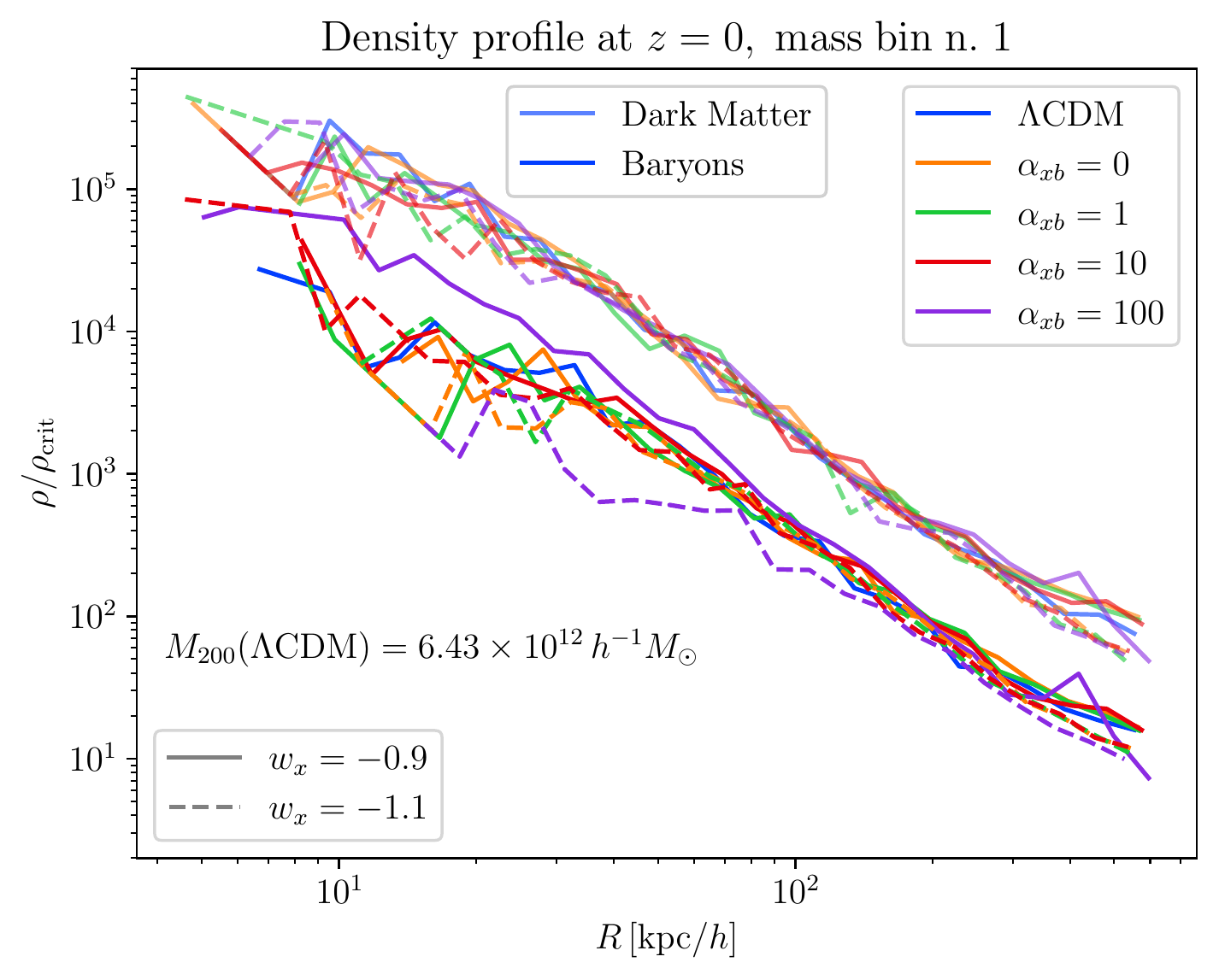} \,\,\,\,\,\,\,\, 
\includegraphics[width=2.8in]{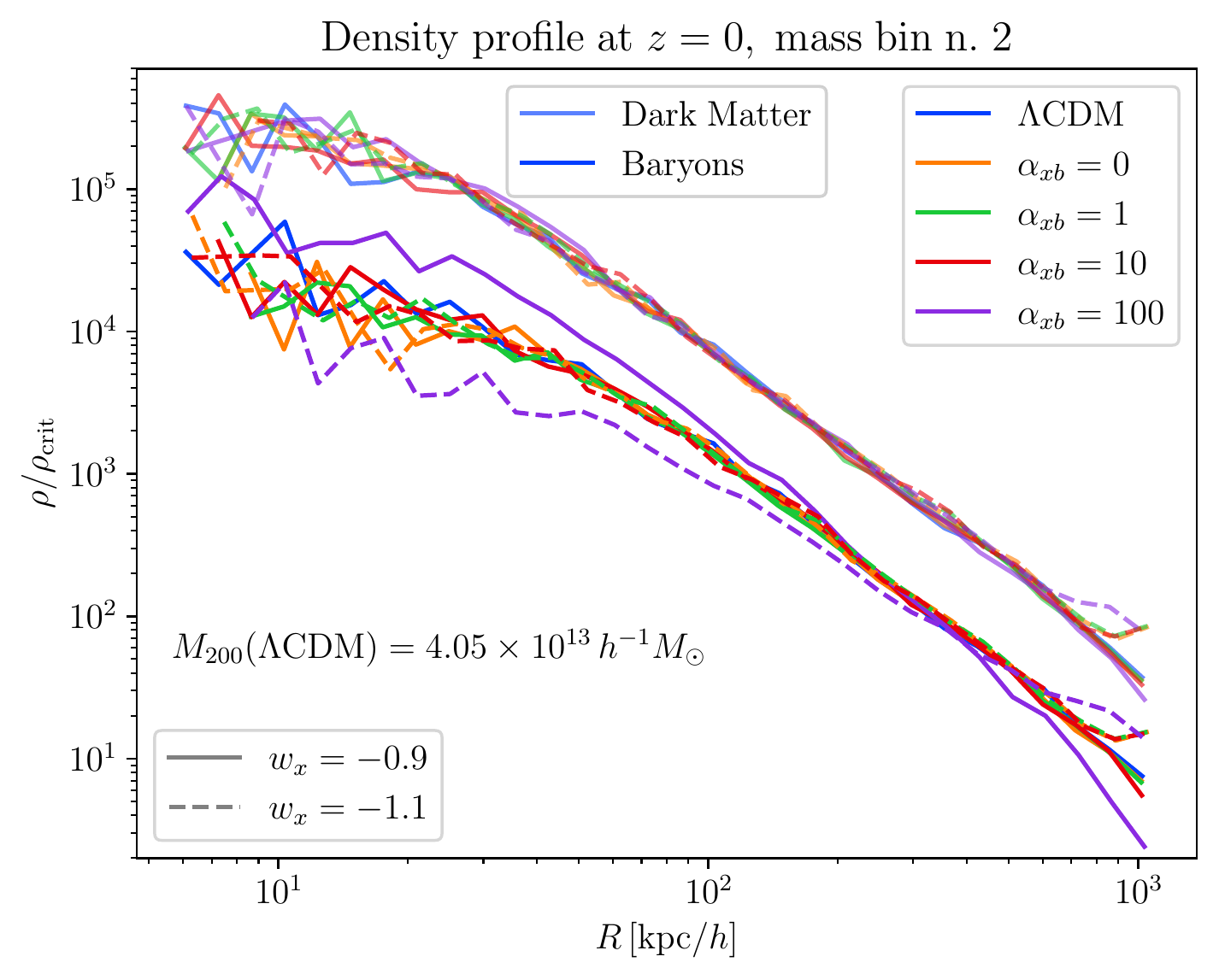} \\
\includegraphics[width=2.8in]{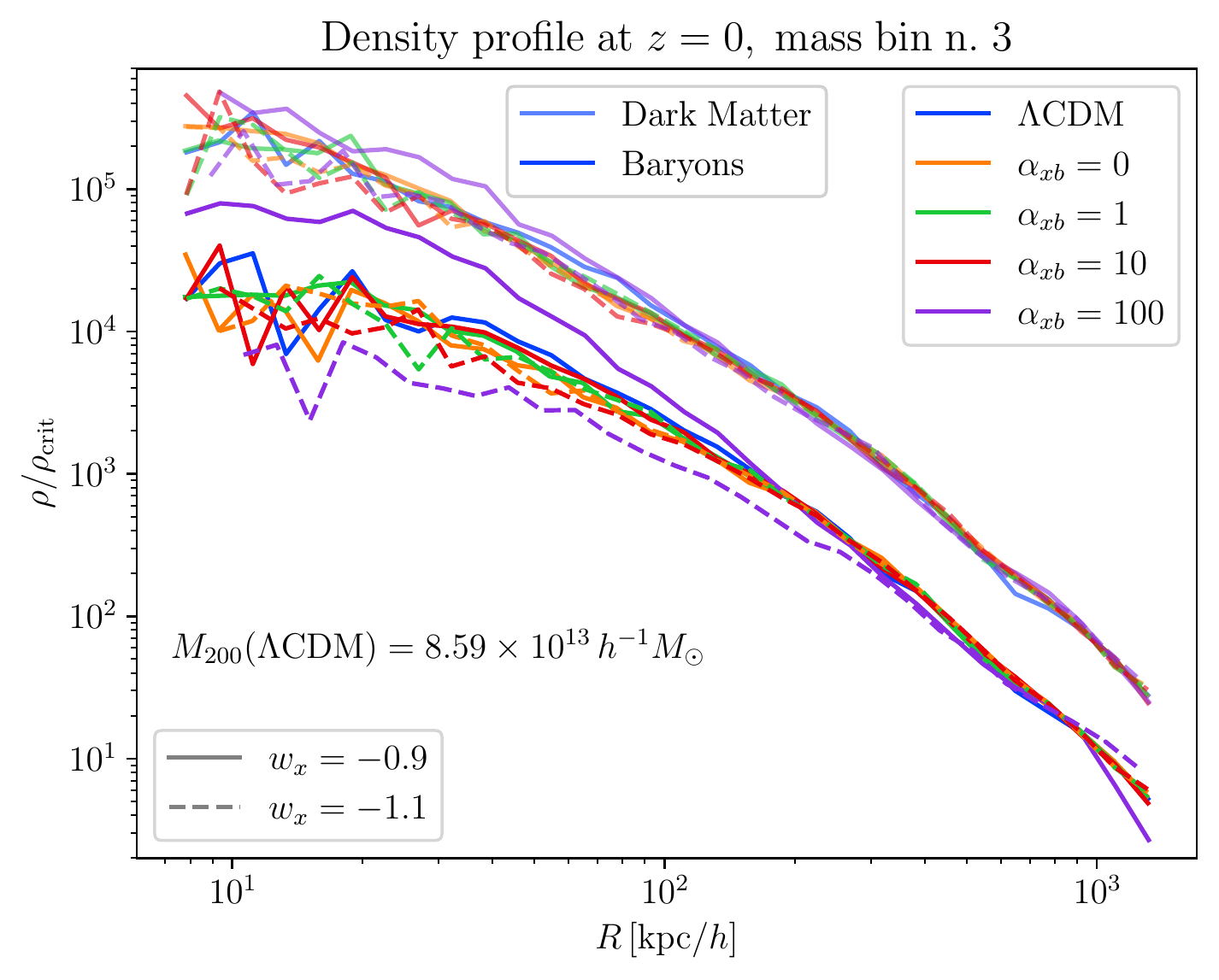} \,\,\,\,\,\,\,\, 
\includegraphics[width=2.8in]{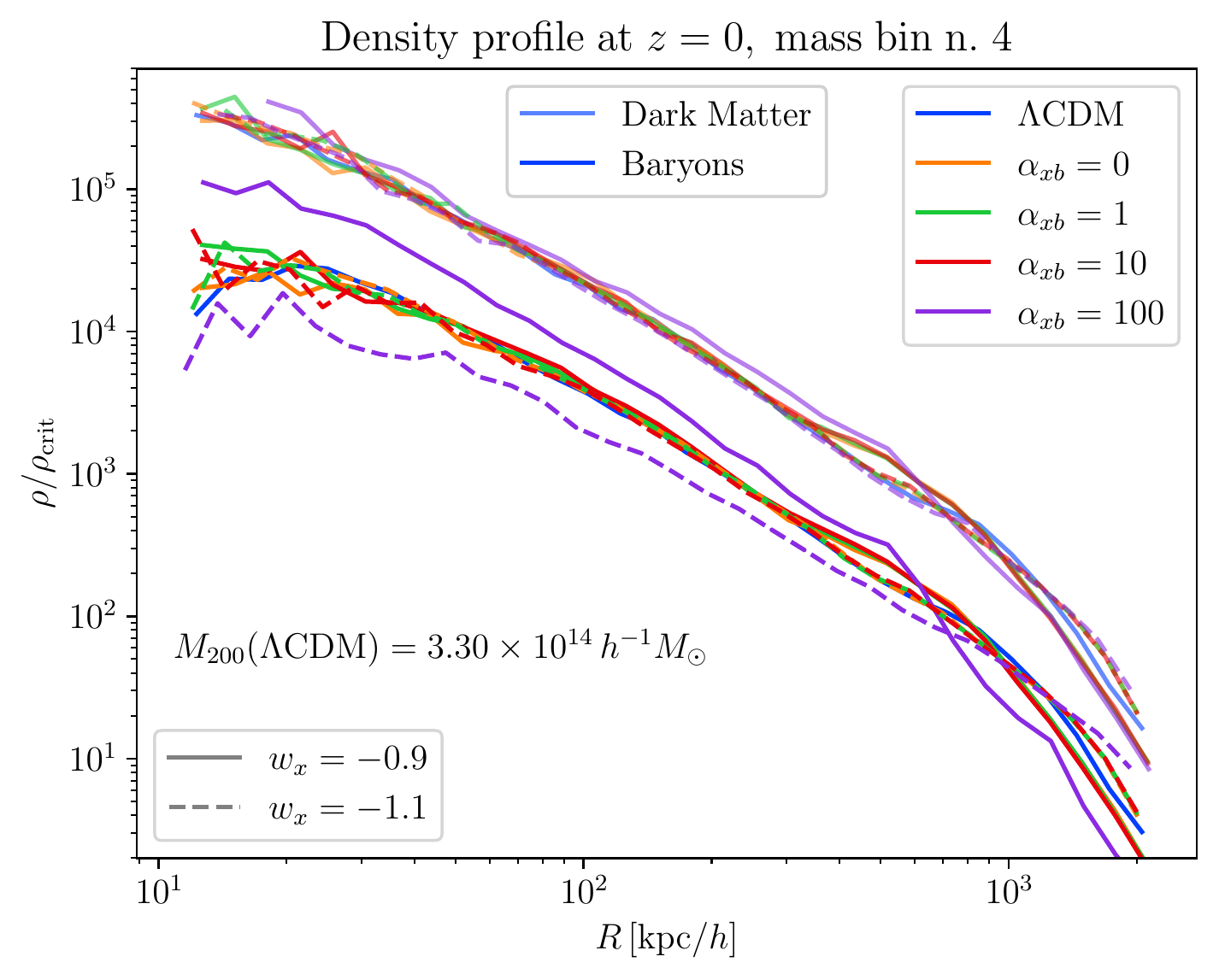}
\caption{Radial halo density profiles at $z=0$ extracted from our simulations within the reference $\Lambda$CDM model (blue curves), and within models featuring dark energy-baryon scattering with Thomson ratio $\alpha_{xb}$ determined by the color coding, whereas the dark energy equation of state is given by $w_x=-0.9$ (solid curves) or $w_x=-1.1$ (dashed curves): in particular, the orange curves correspond to the reference no-scattering $w$CDM models. At a fixed color, we show the density profiles of both the dark matter (shaded curves) and baryon (bright curves) components. Each of the four panels refers to a single halo randomly selected from the four mass bins discussed in the main text (out of a total of 80 halos across all four mass bins).}
\label{fig:halo_profile}
\end{center}
\end{figure*}

\begin{figure*}
\begin{center}
\includegraphics[width=2.8in]{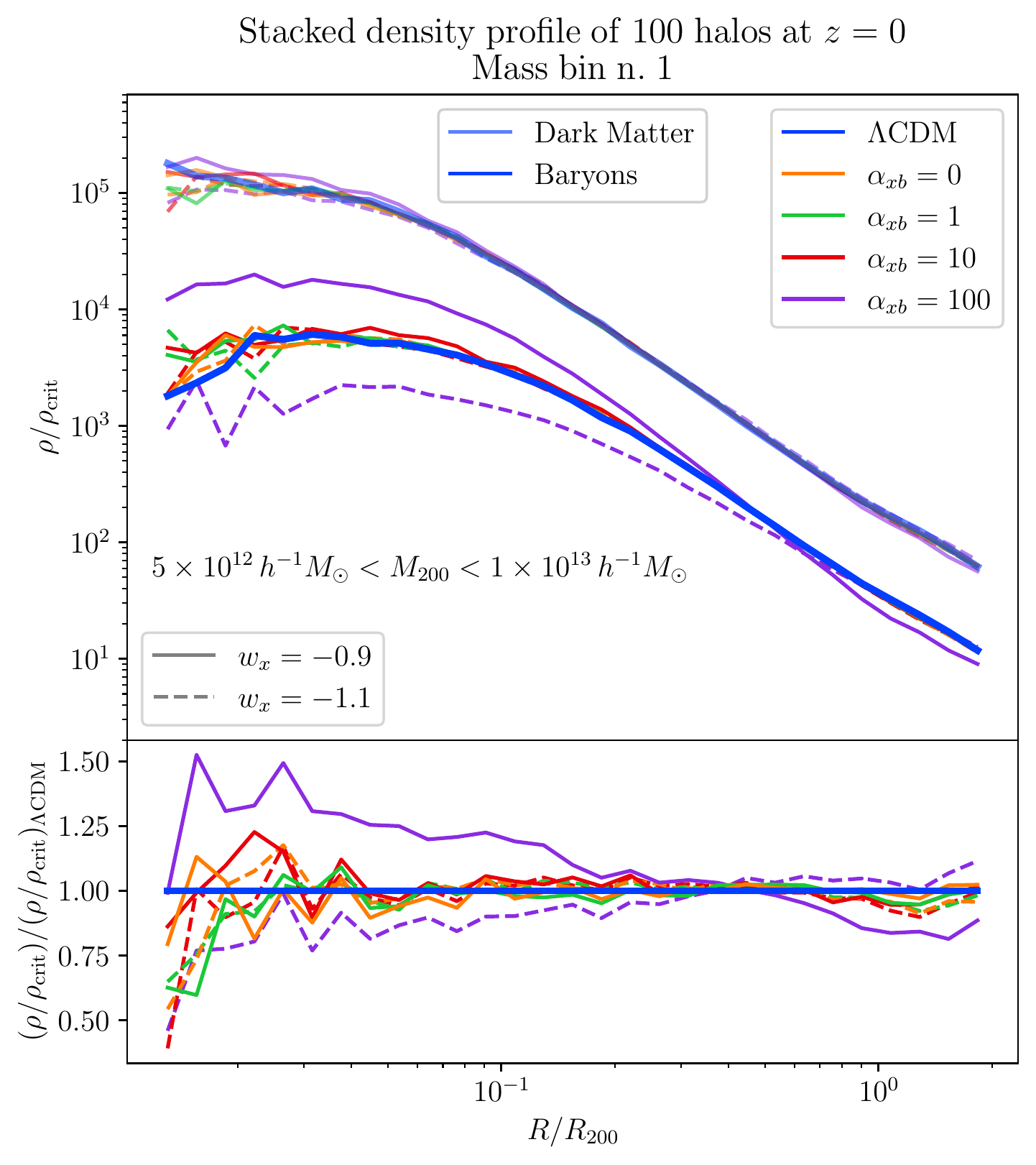} \,\,\,\,\,\,\,\, 
\includegraphics[width=2.8in]{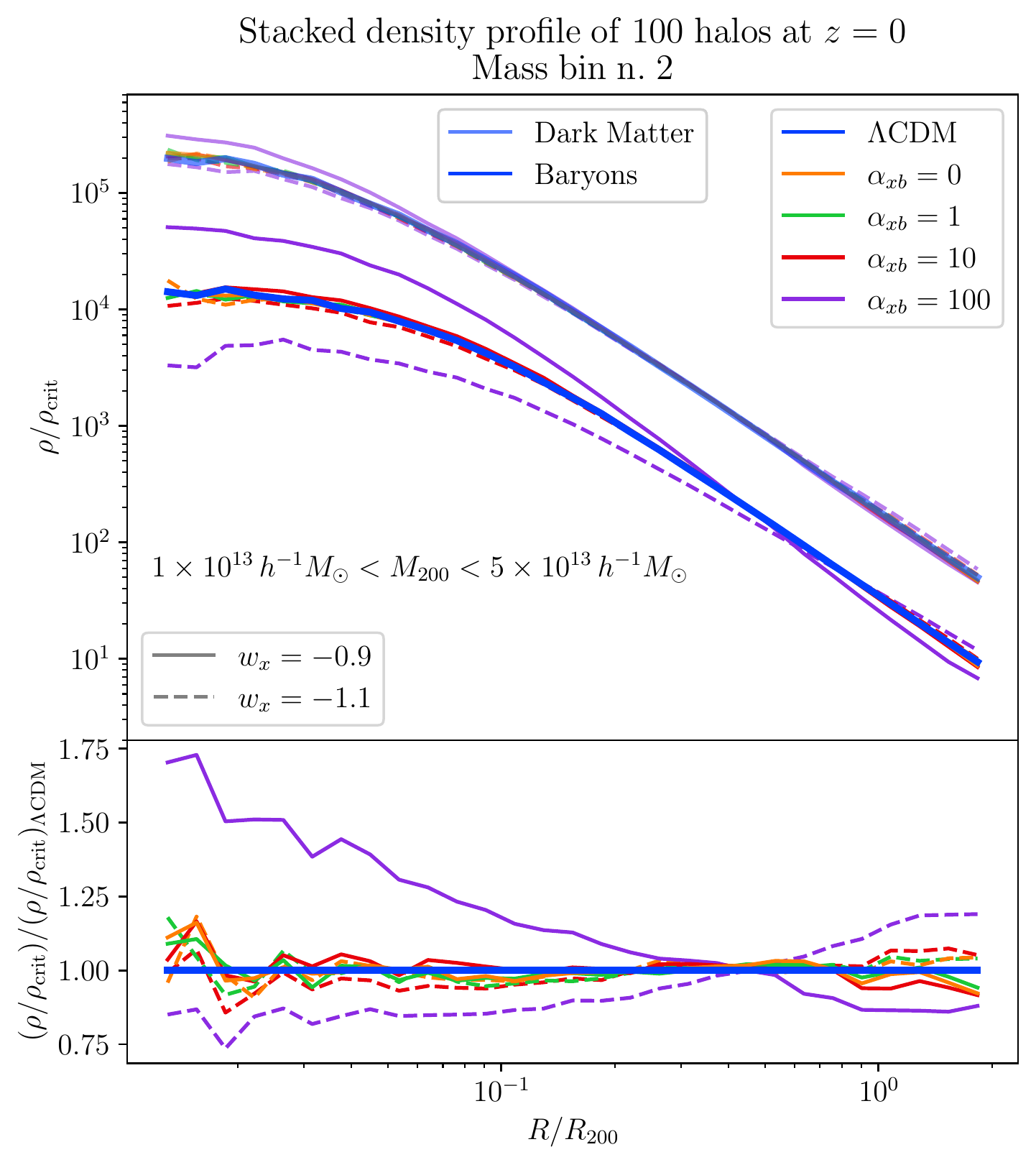} \\
\includegraphics[width=2.8in]{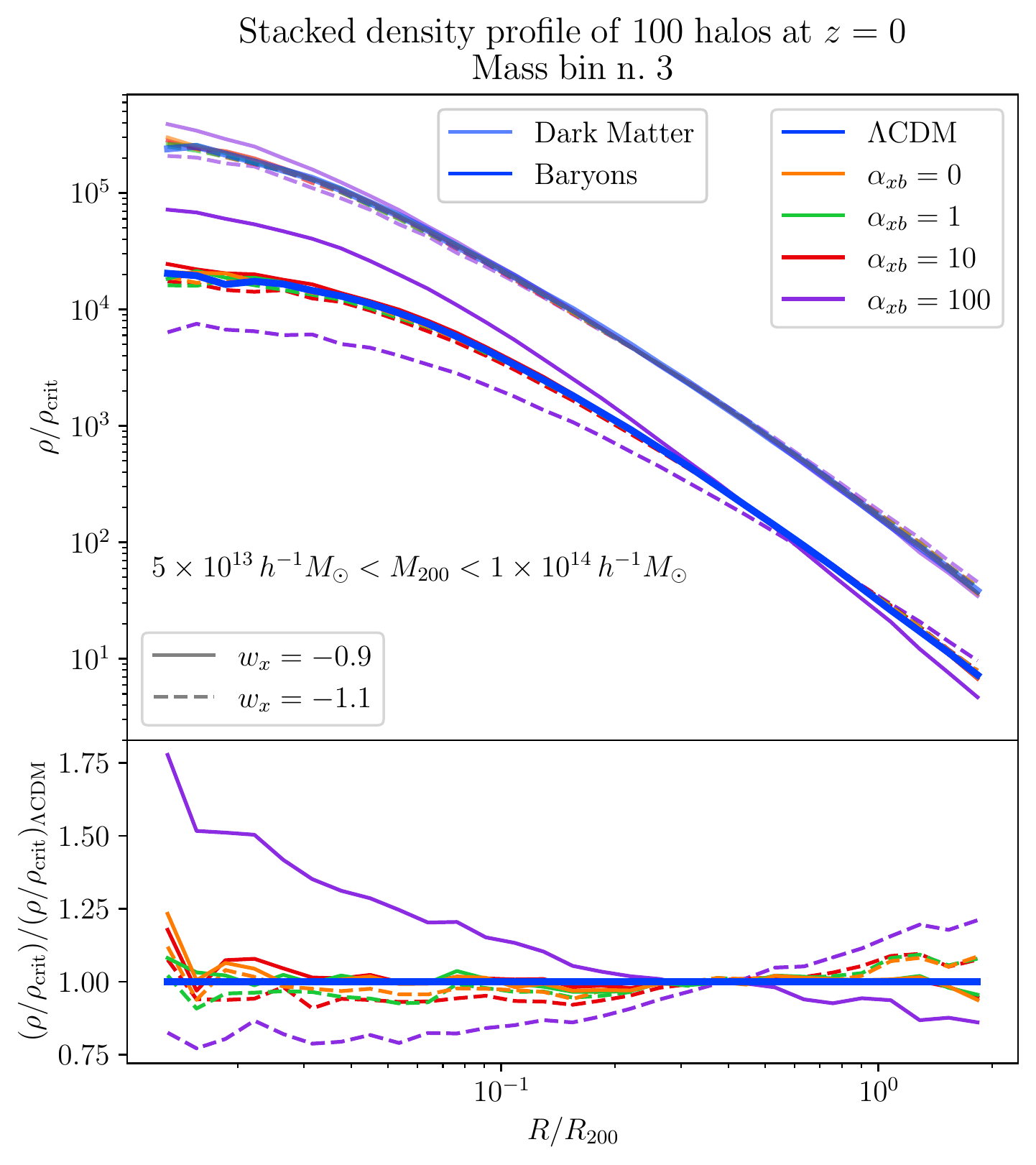} \,\,\,\,\,\,\,\, 
\includegraphics[width=2.8in]{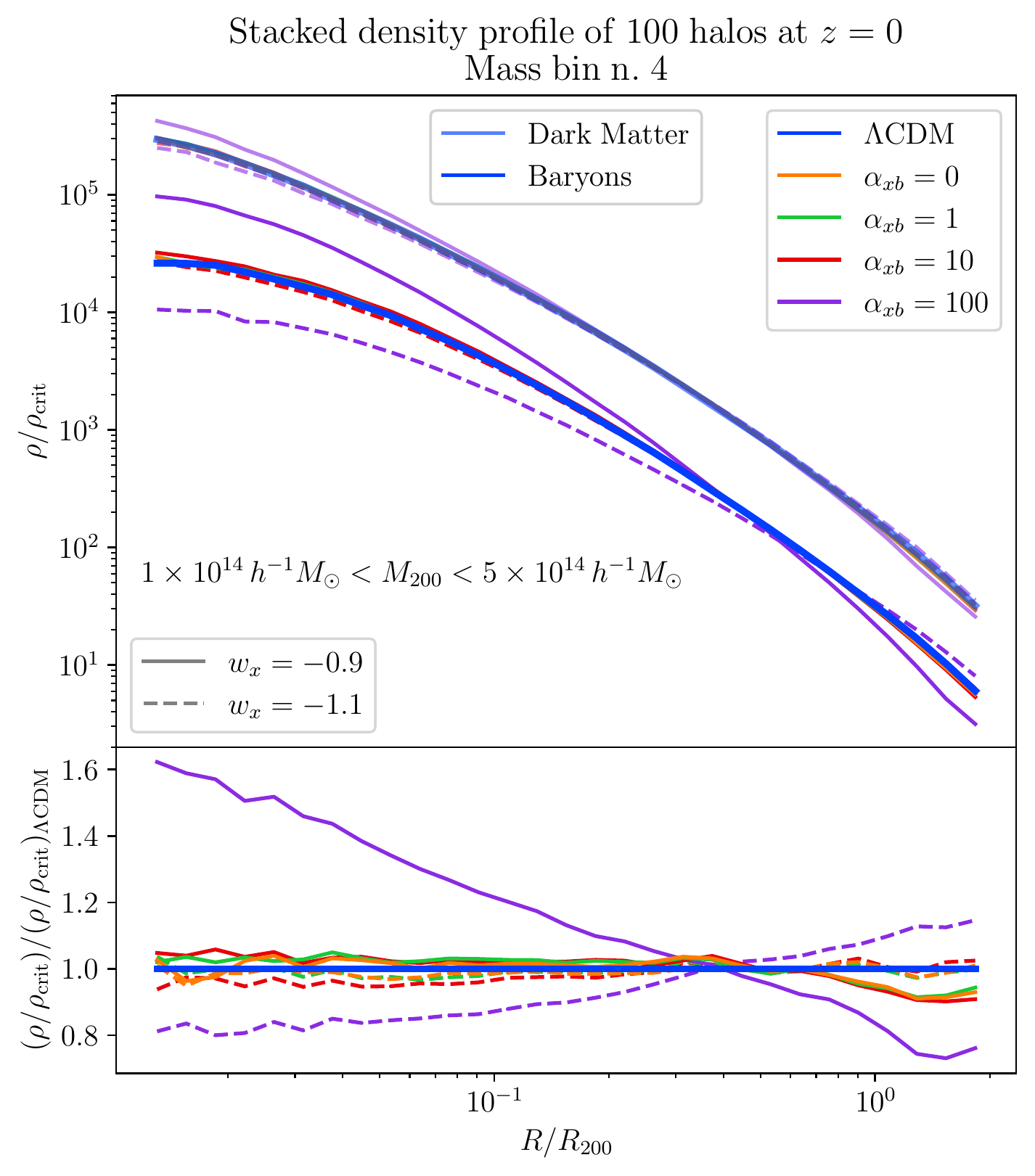}
\caption{For each of the four panels, the upper sub-panels are as in Fig.~\ref{fig:halo_profile}, but this time stacking and averaging the radial density profiles of 100 halos within each of the four mass bins discussed in the main text. The lower sub-panels show the ratios of the total matter density profiles (baryons and cold dark matter) in models with dark energy-baryon scattering (with the same color coding for different values of $\alpha_{xb}$ as in the upper sub-panels) relative to our $\Lambda$CDM simulation.}
\label{fig:halo_stacked_profile}
\end{center}
\end{figure*}

\subsection{Halo profiles}
\label{subsec:haloprofiles}

We now move our investigation to the impact of DE-baryon scattering on the structural properties of halos, focusing on their radial density profiles. To compute these, we begin from the halo catalogues generated by the \texttt{SUBFIND} routine, and select the chosen simulation snapshot. From the snapshot file, we look for the particles making up a given halo. The position of the particle with minimum potential is assumed to be the center of the halo, and spherical shells of logarithmically spaced thickness are considered. We compute the radial density profile by counting the number of particles lying within a given shell, and dividing this quantity by the shell volume. For the purpose of providing a more fair comparison between halos of different virial mass/virial radius, we report the profiles as a function of $R/R_{200}$. We repeat this procedure for both DM and baryonic particles, and for all desired halos. Using the above prescription, for each simulation we compute the radial density profiles for 100 halos within the following four mass bins:
\begin{itemize}
    \item Mass bin 1: $5\times10^{12}\,h^{-1}M_{\odot} < M_{200} < 10^{13}\,h^{-1}M_{\odot}$
    \item Mass bin 2: $10^{13}\,h^{-1}M_{\odot} < M_{200} < 5\times10^{13}\,h^{-1}M_{\odot}$
    \item Mass bin 3: $5\times10^{13}\,h^{-1}M_{\odot} < M_{200} < 10^{14}\,h^{-1}M_{\odot}$
    \item Mass bin 4: $10^{14}\,h^{-1}M_{\odot} < M_{200} < 5\times10^{14}\,h^{-1}M_{\odot}$
\end{itemize}
Subsequently, we consider two different methods to assess the effects of DE-baryon scattering on the density profiles of halos. The first method is effectively a ``direct comparison'', whereas the second method corresponds to a stacked analysis.

To implement the direct comparison method, we work across different simulations and identify objects which can be considered the same halo. We then directly compare their density profiles. Two or more objects identified across different simulations are considered to be the same halo if the following criteria are met:
\begin{itemize}
\item the values of their virial mass and virial radius, $M_{200}$ and $R_{200}$, must not deviate by more than $10\%$;
\item the centers of their structures, defined as the positions of the particles with minimum potential, must not be displaced by more than $80\%$ of $\bar{R}_{200}$, with $\bar{R}_{200}$ being the average value of $R_{200}$ computed across all objects satisfying the first criterion above.
\end{itemize}
Using the above criteria, we detect 80 object in the range $5\times10^{12}\,h^{-1}M_{\odot} < M_{200} < 5\times10^{14}\,h^{-1}M_{\odot}$ which can be considered as being the same halo across different simulations.

In Fig.~\ref{fig:halo_profile} we show four radial density profiles of both DM and baryons at $z=0$, computed from our samples identified with the above criteria. Each panel corresponds to a different mass bin, from which we randomly select a single halo. We note that as the halo mass increases, the noise in the radial density profiles computed from our simulations decreases. The reason is simply that more massive halos are composed of a larger number of particles.

From Fig.~\ref{fig:halo_profile}, we notice in the first place that the radial density profiles computed within the reference $\Lambda$CDM model are consistent with the Navarro-Frenk-White profile~\citep[NFW,][]{Navarro:1995iw}. Next, despite the noise (particularly in the lowest mass bins), we can still observe the effects of DE-baryon scattering, which go in opposite directions depending on whether scattering occurs with a quintessence-like or phantom DE component. Focusing on the quintessence-like case ($w_x=-0.9$), we see that the effect of scattering is to increase the inner density of baryons at all radii, with the effect increasing slightly as we move towards the inner region of the halos, where scattering flattens the density profile making it slightly more cored. We find that the effects of scattering can be rather significant for the most extreme cases, i.e.\ the Q100 and P100 simulations with $\alpha_{xb}=100$, where we observe an increase/decrease in the baryon density by up to factors of $\approx 2-3$. On the other hand, we do not observe significant changes in the DM density profile, even for the $\alpha_{xb}=100$ case. The reason is once more that the effects of DE scattering are only indirectly transmitted to the DM component through gravitational interactions with baryons, as the latter are the only component directly feeling the effects of scattering.

To reduce the noise level and better isolate the effects of DE-baryon scattering on the density profiles of halos, we also adopt a stacking approach. For each mass bin under consideration, we produce a stacked profile at $z=0$ by considering 100 halos with different values of $R_{200}$, and averaging the values of their density profiles as a function of $R/R_{200}$. This allows us to fairly compare halos with different values of $R_{200}$.

The results of this stacking procedure are shown in Fig.~\ref{fig:halo_stacked_profile}. By inspecting the baryon profiles, we confirm the strong increase [decrease] in the inner density for all masses observed earlier through the direct comparison approach, at least for the $\alpha_{xb}=100$ case, where we find an increase of up to factors of $\approx 2-3$. The decrease in noise also allows us to isolate tinier deviations (well below the noise level of the direct comparison method) for the $\alpha_{xb}=10$ case, at the $\approx 10-20\%$ level. On the other hand, the effects of DE-baryon scattering for the $\alpha_{xb}=1$ case are too small to be observed, even with the reduced noise level, and we therefore expect these to be equally challenging to detect in real observations.

We also confirm that the effects on the DM density profile are small, even for the $\alpha_{xb}=100$ case, where the relative deviations are $\lesssim 10\%$. The lower sub-panels of Fig.~\ref{fig:halo_stacked_profile} show the ratios of the total matter density profiles (baryons and CDM) in our simulations relative to $\Lambda$CDM. We see that the deviations in the total matter density profile are also quite significant, particularly when considering scattering between baryons and a quintessence-like DE component. Whether it is the baryon or total matter density profile which is the most relevant observable depends on the observations one is targeting: while the former is relevant for instance for X-ray surveys or probes of the Sunyaev-Zeldovich (SZ) effect, in particular the kinetic and thermal SZ (kSZ and tSZ) effects, the latter is relevant for observations such as weak lensing.

Our stacking approach also allows us to detect another interesting feature, which was previously not visible given the significantly higher noise level. At large radii, the effect of DE-baryon scattering displays an opposite trend compared to that observed at inner radii. That is, scattering with a quintessence-like [phantom] DE component actually decreases [increases] the outer density profile, contrary to the effect we observe in the inner regions. The transition between these two regimes consistently takes place at $\approx 0.5 R_{200}$ within each mass bin.

Our results can again be explained in terms of the role of angular momentum in collapsing structures, which also underlies the effects found in other observables and discussed earlier. Specifically, the dissipation [injection] of kinetic energy from [into] the system caused by scattering between baryons and a quintessence-like [phantom] DE component moves baryons towards the inner [outer] regions of the halos, with the net flow being negligible at $\approx 0.5 R_{200}$, where we observe the transition between the different behaviors in the inner and outer regions reported previously.

\subsection{Halo baryon fraction profiles}
\label{subsec:halobaryonfractionprofiles}

\begin{figure*}
\begin{center}
\includegraphics[width=2.8in]{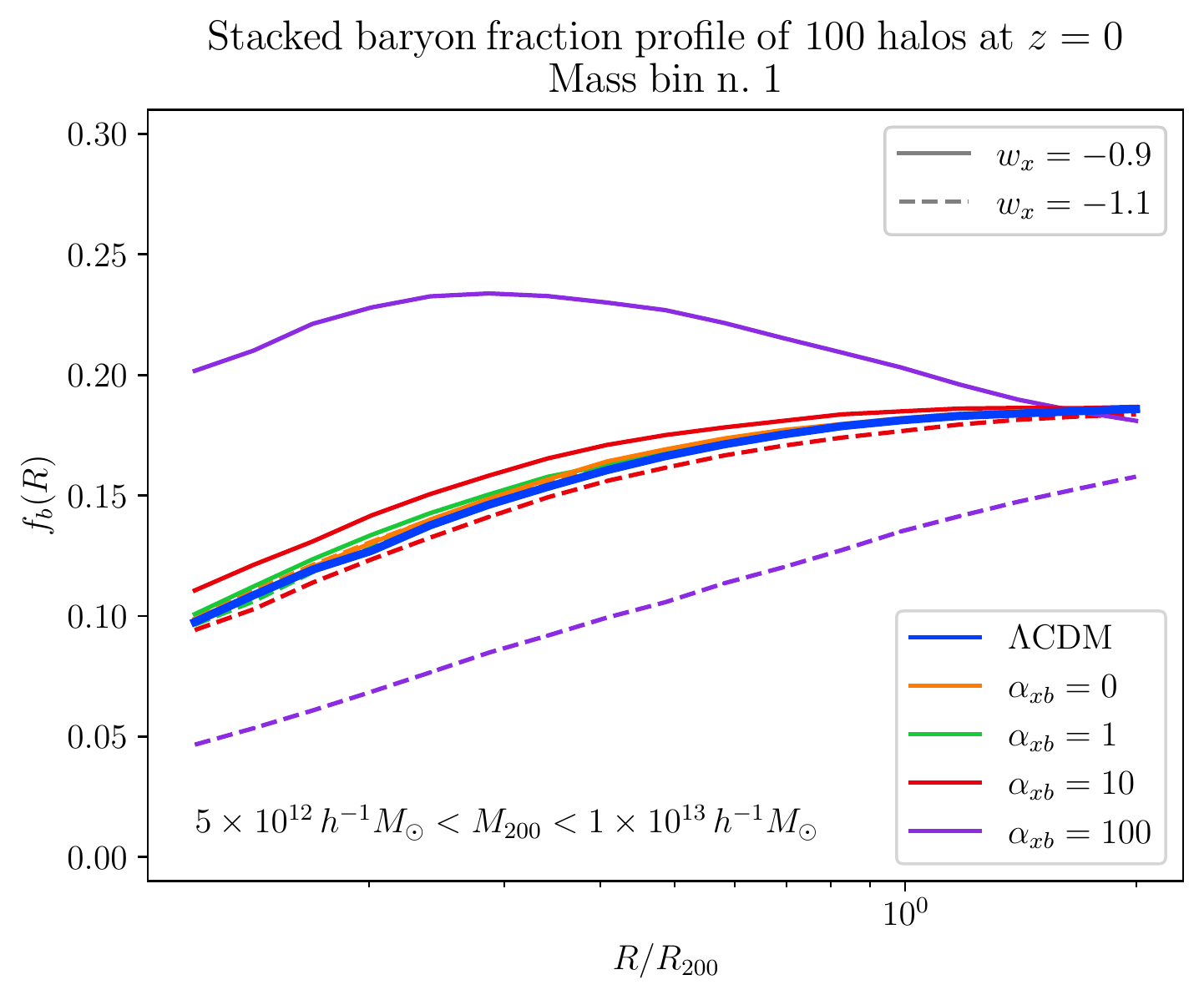} \,\,\,\,\,\,\,\, 
\includegraphics[width=2.8in]{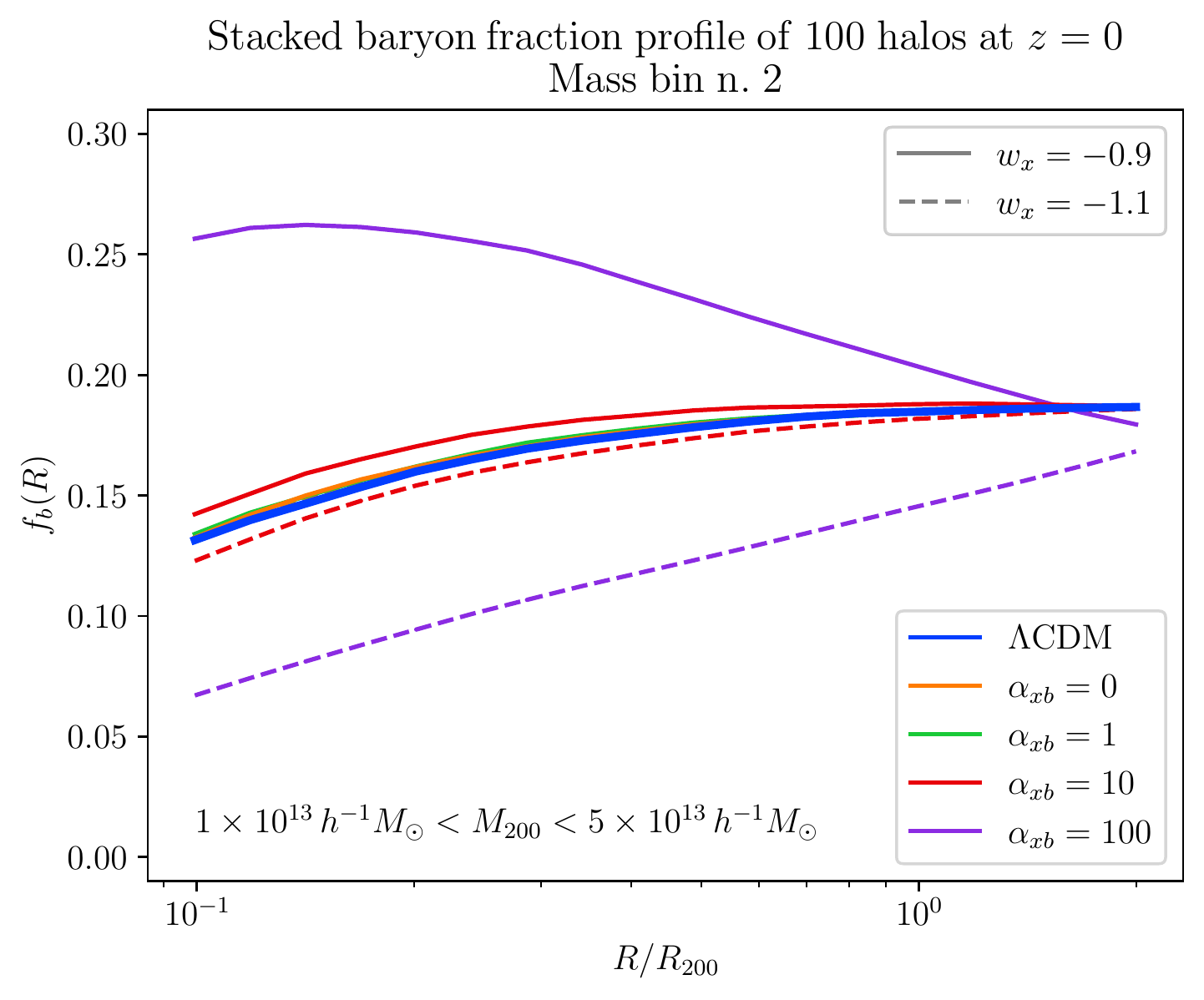} \\
\includegraphics[width=2.8in]{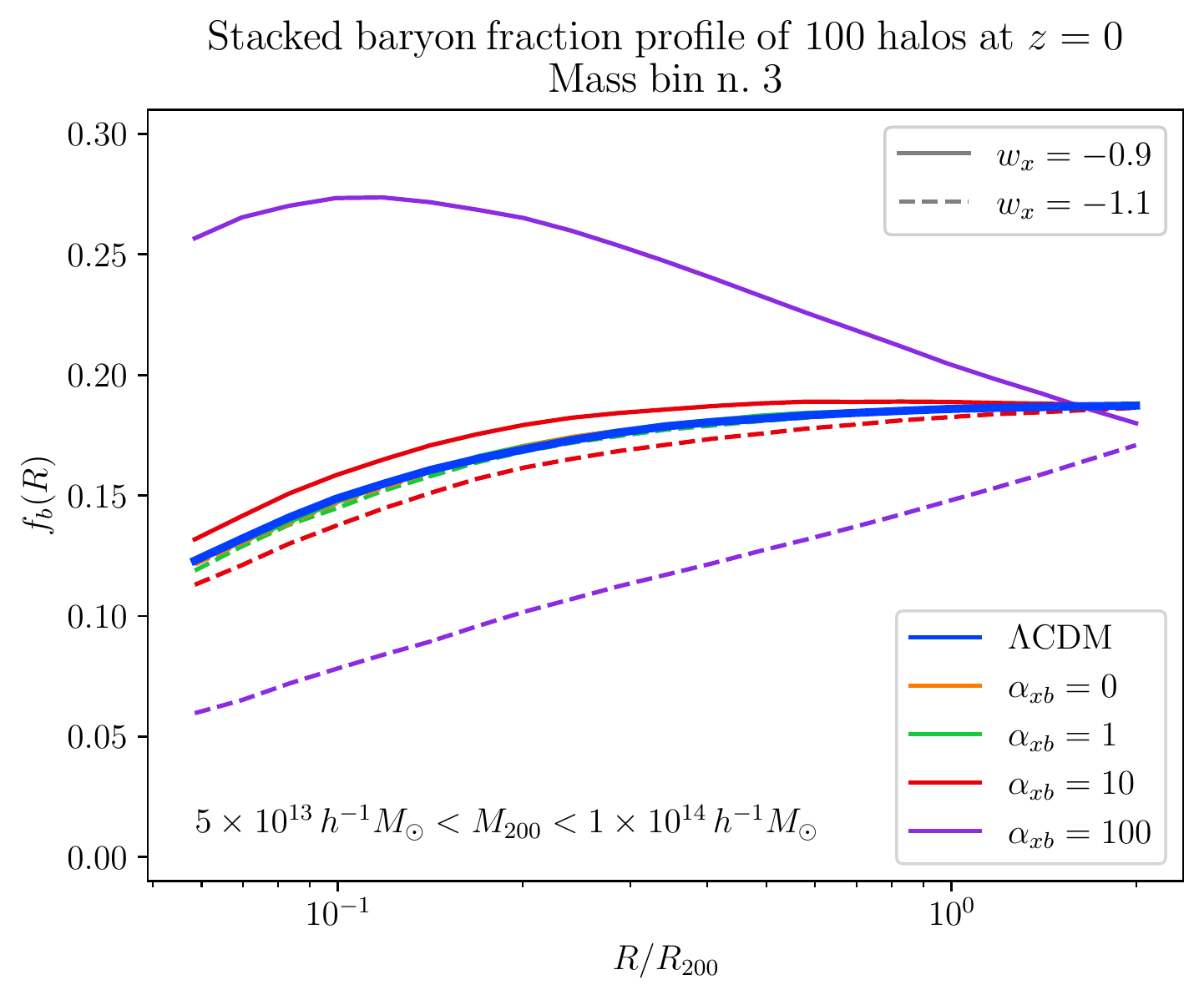} \,\,\,\,\,\,\,\, 
\includegraphics[width=2.8in]{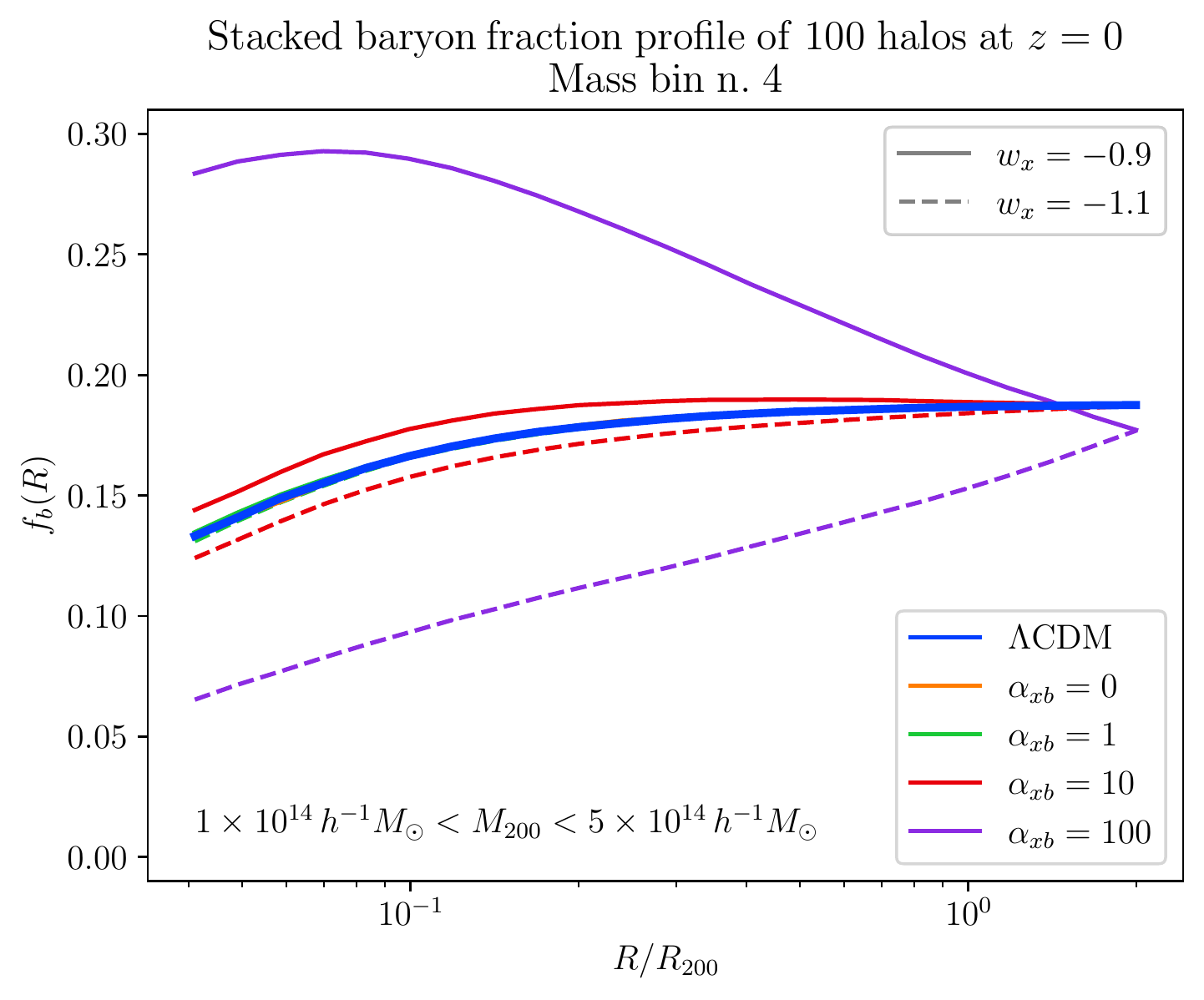}
\caption{As in Fig.~\ref{fig:halo_stacked_profile}, but focusing on the halo baryon fraction profiles. We again proceed by staking the baryon fraction profiles of 100 halos within each of the four mass bins discussed in the main text.}
\label{fig:baryon_fraction_stacked_profile}
\end{center}
\end{figure*}

As a final observable, we consider the baryon fraction profiles of halos. To compute these, we consider ratio between the cumulative mass profiles of baryons and DM. If the baryon and DM density profiles are given by $\rho_b(R)$ and $\rho_c(R)$ respectively, their cumulative mass profiles $M_b(R)$ and $M_c(R)$ are given by the following integrals:
\begin{eqnarray}
M_{b,c}(R) \equiv 4\pi\int_0^R dr\,r^2\rho_{b,c}(r)\,,
\label{eq:cumulativemassprofile}
\end{eqnarray}
which we evaluate via discrete sums over spherical shells. We then define the baryon fraction profile as $f_b(R) \equiv M_b(R)/M_c(R)$.

To understand the effects of DE-baryon scattering, in this case we directly opt for a stacking approach, given that the noise level would otherwise be too high as we have previously seen. In Fig.~\ref{fig:baryon_fraction_stacked_profile} we show the stacked baryon fraction profiles obtained by averaging the baryon fraction profiles as a function of $R/R_{200}$ of 100 halos in each of the four mass bins. In the reference $\Lambda$CDM model, as well as in the reference no-scattering $w$CDM models (Q0 and P0 simulations), we see a similar trend for the innermost regions of the baryon fraction profiles, with $f_b \approx 0.1$ for the least massive halos and $f_b \approx 0.14$ for the most massive ones. After a transition region where $f_b(R)$ grows, the baryon fraction then settles to $\approx 0.19$: this is expected, as $0.19$ approximately corresponds to the ratio $\Omega_b/(\Omega_m-\Omega_b)$, i.e\ the expected cosmic baryon fraction at $z=0$ given the choice of cosmological parameters we have adopted in our simulations. The systematically lower baryon fraction in the inner regions could be connected to dynamical friction, which acts preferentially on CDM particles as they are more massive than baryonic ones, and/or to numerical heating. We defer a more complete investigation of this issue to future work.

Let us now inspect the effects of DE-baryon scattering. For the $\alpha_{xb}=10$ case we observe a systematic $\approx 5\%$ enhancement [suppression] of the baryon fraction profile for $w_x=-0.9$ [$w_x=-1.1$], at least in the inner regions. However, as $R$ approaches $R_{200}$, these effects are damped and $f_b$ converges to the cosmic baryon fraction. The relative deviations in the baryon fraction profiles observed for the $\alpha_{xb}=10$ case appear to be approximately scale-independent.

When we move on to the more extreme $\alpha_{xb}=100$ case, we find that the baryon fraction profiles are significantly distorted, i.e.\ the relative deviations with respect to the reference cases are highly scale-dependent. For the $w_x=-0.9$ case (Q100 simulation), we notice a strong enhancement of the baryon fraction profile in the inner regions, leading to relative deviations of $\approx {\cal O}(1)$. Subsequently, the baryon fraction profile exhibits a relatively steep decrease, reaching $f_b \approx 0.19$ at $R\approx 1.5R_{200}$, and even lower values at larger radii. For the $w_x=-1.1$ case (P100 simulation), we observe a smaller overall suppression of $\approx 50\%$ in the inner regions. The baryon fraction profile then exhibits a steady growth, approximately mirroring the decrease observed in the $w_x=-0.9$ case. Finally, we also observe that even at $R\approx 2R_{200}$, $f_b$ is slightly lower compared to its value in the reference models.

The physical explanation for these results can again be traced to the role of angular momentum in collapsing structures. In fact, the friction [drag] force related to scattering between baryons and a quintessence-like [phantom] DE component supports [opposes] the loss of angular momentum, which in turn helps [hinders] the gravitational collapse of baryons in DM halos. Therefore, the baryon fraction profile will correspondingly increase [decrease] in each of the two cases. Compared to the effects on the density profiles observed earlier, the effects on the baryon fraction profiles are enhanced due to appearance of the DM mass profile in the denominator, given that the DM components of halos are significantly less affected by DE-baryon scattering as discussed earlier. It is important to note that the significant modifications to the baryon fraction profiles brought about by DE-baryon scattering in the most extreme cases ($\alpha_{xb}=100$) are expected to leave an equally significant impact on several non-adiabatic astrophysical processes, including radiative cooling, star and galaxy formation, SNe and AGN feedback, and so on. To properly study these scenarios would require running appropriate hydrodynamical simulations, wherein the usual $\Lambda$CDM-driven implementation of these non-adiabatic processes might no longer be correct, an extremely important issue whose investigation however we defer to future work.

In summary, we have found that DE-baryon scattering leaves significant imprints in the structural properties of halos, such as their density and baryon fraction profiles. As with the other observables we considered earlier, these imprints are related to the role of angular momentum in collapsing structures. As we shall discuss soon, these signatures may potentially be detectable with next-generation cosmological and astrophysical observations even for $\alpha_{xb}$ as small as $\approx {\cal O}(1)$.

\section{Discussion}
\label{sec:discussion}

Our simulations have provided a first glimpse into the signatures of scattering between DE and baryons in the non-linear regime of structure formation. We have shown that our model for scattering, while phenomenological, displays a rich array of observational signatures, the majority of which are ultimately tied to the role of angular momentum in collapsing structures, whose virial equilibrium can be significantly altered by DE-baryon scattering. These signatures could possibly be within the reach of future cosmological and astrophysical observations.

Among all the observables considered, cosmological observations of the clustering of the LSS appear among the least promising. At the level of matter power spectrum, the effects of DE-baryon scattering only become prominent at highly non-linear scales ($k \sim 10\,h{\rm Mpc}^{-1}$). The real challenge in this context is that of properly modelling the underlying theoretical power spectrum, including complications associated to non-linearities and astrophysical processes~\citep[see e.g.][]{Brun:2013yva,Schaye:2014tpa,McCarthy:2016mry}. Recall that our simulations do not include non-adiabatic processes such as radiative cooling, star and galaxy formation, and SNe and AGN feedback. As these processes are driven to an important extent by the baryonic component in galaxies, we expect them to be affected by scattering with DE: therefore, including non-adiabatic processes in our simulations may be highly non-trivial, and it is certainly unclear whether the $\Lambda$CDM-oriented implementation of such processes adopted in several other studies can be safely carried on to our case. A full investigation of this issue is well beyond the scope of this paper and requires a dedicated follow-up work, which we defer to a future study.

Another issue which is not addressed by our N-body simulations is the fact that we do not directly observe the underlying matter field, but only biased tracers thereof~\citep{Desjacques:2016bnm}. The issue of LSS tracer bias further complicates the search for signatures of DE-baryon scattering in the non-linear regime.~\footnote{On a more positive note we also point out that, in principle, DE-baryon scattering could affect the bias of LSS tracers, in a way similar to massive neutrinos~\citep[see e.g.][]{Castorina:2013wga,LoVerde:2014rxa,LoVerde:2014pxa,Raccanelli:2017kht,Munoz:2018ajr,Vagnozzi:2018pwo,Valcin:2019fxe}. This could open a new pathway towards searching for signatures of DE-baryon scattering in cosmological observations.} Nevertheless, assuming that future LSS surveys such as Euclid~\citep{EUCLID:2011zbd} and DESI~\citep{2016arXiv161100036D} will be able to reliably model the power spectra of biased LSS tracers in the highly non-linear regime, it will be possible to search for signs of DE-baryon scattering with strength $\alpha_{xb} \sim {\cal O}(100)$ if the modelling is precise to the ${\cal O}(10\%)$ level, and $\alpha_{xb} \sim {\cal O}(10)$ if the modelling is precise to the ${\cal O}(1\%)$ level. Note, however, that an approach introducing a theoretical error encoding our ignorance of higher-order non-linear corrections~\citep[see e.g.][]{Sprenger:2018tdb,Brinckmann:2018owf,Chudaykin:2019ock,Chudaykin:2020hbf,Steele:2020tak} can potentially wash out a significant amount of information contained in non-linear modes: if a very conservative approach such as that presented in~\cite{Sprenger:2018tdb} were adopted, in the best case one would only be able to look for signatures of DE-baryon scattering with strength $\alpha_{xb} \gtrsim {\cal O}(100)$, as the signatures of scattering with weaker strength would be below the theoretical error budget. Overall it is clear that, in continuing our exciting program for the search of cosmological and astrophysical signatures DE-baryon scattering, a more complete understanding of the effects thereof on the non-linear clustering of tracers of the LSS including non-adiabatic processes is of paramount importance, and is a priority for follow-up work.

On the other hand, the signatures of DE-baryon scattering on the abundance and structures of halos are significantly more promising, observationally speaking. For instance, scattering with strength $\alpha_{xb} \sim {\cal O}(100)$ leads to ${\cal O}(10\%)$ changes in the abundance of halos at all masses, whereas the changes are of ${\cal O}(1\%)$ for $\alpha_{xb} \sim {\cal O}(10)$. However, the halo mass function is not a directly observable quantity, as halo masses are difficult to measure. Strictly speaking, one should therefore connect the halo mass function to more readily observable quantities, such as the luminosity function, baryon mass function, or rotational velocity function of LSS tracers such as galaxies: see for instance~\cite{Zwaan:2009dz,Papastergis:2012wh,Bouwens:2014fua,Klypin:2014ira,Tortorelli:2020ovx} for examples of these measurements.

A complete investigation in this sense requires extending our simulations to fully understand how the effects we observed on halos propagate to luminous tracers of the LSS distribution such as galaxies, which in turn requires assessing the impact (if any) of DE-baryon scattering on halo occupation distribution models. This is a task which goes significantly beyond the scope of the present work, but which once more is a priority for follow-up work. Finally, it is worth mentioning the existence of proposals for directly measuring the halo mass function using combinations of weak lensing, galaxy cluster counts, galaxy cluster power spectra, and lensed Type Ia Supernovae data~\citep{Castro:2016jmw}, as well as using sub-mm magnification bias~\citep{Cueli:2021dai}: should applying these methods ultimately prove feasible on real data, they would undoubtedly provide a very interesting pathway towards directly probing signatures of DE-baryon scattering on the halo mass function.

Our simulations show that that the signatures of DE-baryon scattering on the structures of halos (density profiles and baryon fraction profiles) are the most promising from the observational point of view. We have found that scattering with strength $\alpha_{xb} \sim {\cal O}(100)$ leads to ${\cal O}(100\%)$ or larger changes in the density profiles of baryons and the baryon fraction profiles, whereas these same observables are altered by ${\cal O}(10\%)$ for scattering with strength $\alpha_{xb} \sim {\cal O}(10)$. At present, these observables can potentially be measured to the $\approx 10\%$ level (or even better) by means of a wide array of methods, which include but are not limited to weak and strong lensing~\citep{2005A&A...442..413M,Vegetti:2014wza,Umetsu:2015baa}, X-ray surface brightness or temperature~\citep{Nagai:2006sz}, the line-of-sight velocity dispersion of stars and satellite galaxies~\citep{Battaglia:2005rj,More:2010sf,Yildirim:2015saq}, as well as the SZ effect~\citep{Battaglia:2017neq,AtacamaCosmologyTelescope:2020wtv,2021arXiv211002228S}.

As shown in the forecasts of~\cite{Battaglia:2017neq}, a combination of kSZ and tSZ measurements should enable measurements of the baryonic density profiles of halos to the $1\%$ level with near-future CMB and LSS surveys such as AdvACT~\citep{Henderson:2015nzj}, SPT-3G~\citep{SPT-3G:2014dbx}, the Simons Observatory~\citep{SimonsObservatory:2018koc,SimonsObservatory:2019qwx}, CMB-S4~\citep{2016arXiv161002743A}, and DESI~\citep{2016arXiv161100036D}, and including weak lensing measurements can be used to provide tighter constraints on the baryon fraction profiles~\citep{2021arXiv211002228S}. If near-future cosmological probes will be able to reliably measure the baryon density and baryon fraction profiles of halos down to the $\%$ level (or better), it could in principle be possible to search for signs of DE-baryon scattering with strength as small as $\alpha_{xb} \sim {\cal O}(1)$, which otherwise appears to be prohibitive with all the other probes we have considered (such as LSS clustering and halo mass function).

Overall, it therefore appears that prospects for probing DE-baryon scattering with strength $\alpha_{xb} \lesssim {\cal O}(100)$ from near-future cosmological and, especially, astrophysical observables, are very bright. In fact, the tiny effects observed in the linear regime by~\cite{Vagnozzi:2019kvw} are significantly enhanced in the non-linear regime. This is of course contingent upon the ability to reliably model the observables in question at least to the ${\cal O}(10\%)$ and possibly ${\cal O}(1\%)$ level, which in some cases (e.g. non-linear LSS clustering) is expected to be particularly challenging. In addition, we note that DE-baryon scattering with strength as large as $\alpha_{xb} \sim {\cal O}(1-10)$ may be motivated by the XENON1T direct detection excess~\citep{XENON:2020rca}, in light of the recent interpretation provided by one of us in~\cite{Vagnozzi:2021quy}, envisaging scattering between (chameleon-screened) DE and electrons.

Before closing, we also outline a number of other follow-up directions and observables it might be interesting to consider beyond those discussed in this work. The fact that DE scatters with baryons but not with DM is expected to lead to a ``baryon bias'' effect, where baryons feel an extra drag or pull which will result in their being slower or faster than the corresponding DM halos. This effect can potentially be searched for in astrophysical observations by exploiting the motion of tidally disrupted stellar streams~\citep{Amendola:2001rc,Kesden:2006vz,Kesden:2006zb}. However, the baryon bias effect may be particularly evident in systems of merging cluster such as the Bullet Cluster~\citep{Randall:2008ppe}, or more generally in Bullet-like systems~\citep{Bradac:2008eu}, displaying a clear segregation between collisional matter (baryons) observed via X-rays and collisionless matter (DM) observed via lensing. A potentially interesting follow-up direction would therefore be to identify Bullet Cluster-like systems in our simulations, re-simulate them with a zoom technique~\citep{Navarro:1994zk}, and study how the offset between the baryonic and DM components is affected by DE-baryon scattering. Moreover, given the important role of angular momentum in collapsing structures in driving the signatures we have observed, it could be worth studying in more detail both the velocity dispersion profiles of halos, as well as their shapes. In addition, as alluded to previously, the bias of LSS tracers might also contain the imprint of DE-baryon scattering, particularly on small scales. Finally, the effects of DE-baryon scattering might be more pronounced in higher-order correlators of the matter field beyond the power spectrum, such as the matter bispectrum and trispectrum~\citep[see e.g.][for recent studies thereof]{Gualdi:2020eag,Gualdi:2021yvq}, whose inclusion could also naturally help break the large-scale degeneracy between $\alpha_{xb}$ and $\sigma_8$: it could therefore be interesting to extract these observables from our simulations and study their response to DE-baryon scattering. We leave a more detailed exploration of these and other observables to follow-up work.

\section{Conclusions and outlook}
\label{sec:conclusions}

The possibility that dark energy (DE) might enjoy non-gravitational interactions is a well-motivated one~\citep{Simpson:2010vh}. Analogously to searches for non-gravitational interactions of dark matter (DM), which are currently the state-of-the-art in experimental searches for DM, looking for non-gravitational signatures of DE can help deliver significant insight into the fundamental nature of the mysterious component driving the accelerated expansion of the Universe. In this work, we have investigated the theoretically well-motivated possibility of scattering between DE and baryons, with the scattering strength quantified by the dimensionless quantity $\alpha_{xb}$, given by the ratio of the DE-baryon scattering cross-section to the Thomson cross-section. As in the case of DM, such a scattering process might open the window towards \textit{direct detection} of DE. The earlier work of~\cite{Vagnozzi:2019kvw} argued that, even in the case of strength $\alpha_{xb} \sim {\cal O}(1)$ or larger, the imprints of DE-baryon scattering on cosmological observables in the linear regime are far too small to be observed in near-future surveys. In this work, we have extended these earlier results to the non-linear regime, and investigated for the first time the signatures of DE-baryon scattering on the non-linear formation of cosmic structures.

We have run a suite of large N-body simulations incorporating the effects of scattering between baryons and a perfect fluid associated with DE (with strength $1 \lesssim \alpha_{xb} \lesssim 100$), and extracted a number of relevant observables related to statistical properties of the large-scale structure and/or the structural properties of collapsed halos, including the non-linear matter power spectrum, halo mass function, halo density profiles, and halo baryon fraction profiles. When compared to reference $\Lambda$CDM or no-scattering $w$CDM simulations, we have found that the deviations in these observables due to DE-baryon scattering in the non-linear regime typically exceed their linear counterparts by a significant amount (in some cases by orders of magnitude), changing sign in the case of the matter power spectrum. Most of the deviations we have observed can be linked to the role of angular momentum in collapsing structures, whose virial equilibrium can be significantly altered by DE-baryon scattering. Of all the observables we have considered, the baryon density profiles and baryon fraction profiles of halos appear to be the most promising ones. A combination of near-future kSZ, tSZ, and weak lensing measurements can in principle measure these observables to the $1\%$ level or better~\citep{Battaglia:2017neq}, potentially allowing us to probe DE-baryon scattering strengths down to the level of $\alpha_{xb} \sim {\cal O}(1)$, which could be motivated by the chameleon-screened DE interpretation of the XENON1T excess~\citep{XENON:2020rca} discussed in~\cite{Vagnozzi:2021quy}.

Our results, which demonstrate that scattering between DE and baryons leads to a very rich phenomenology in the non-linear formation of cosmic structures, are not the final word on the subject, as there are several interesting follow-up avenues. Of paramount importance in continuing our exciting program for the search of cosmological and astrophysical signatures of DE-baryon scattering is a thorough understanding of the impact of non-adiabatic processes (not included in our simulations), and of the impact of DE-baryon scattering on the latter. In addition, we have argued that it is worthwhile to study additional observables beyond those considered here: a particularly relevant example in this sense are Bullet-like systems of merging clusters, which might contain the imprint of DE-baryon scattering in the form of a ``baryon bias''. Finally, another interesting direction could be to perform a robust forecast for the ability of future cosmological and astrophysical surveys to probe DE-baryon scattering: an important step in this sense would be to develop a tractable method to interpolate across computationally expensive N-body simulations, for instance through an emulator approach~\citep[see e.g.][]{Rogers:2018smb,Rogers:2020cup,Mancini:2021lec,Carrilho:2021rqo}. While we set aside these and related issues for follow-up work, our current results demonstrate that prospects for \textit{cosmological/astrophysical direct detection of dark energy} are extremely bright, and that we might well be able to detect the first unambiguous non-gravitational fingerprints of DE over the next decade.

\section*{Data Availability} 
The data underlying this article will be shared upon reasonable request to the corresponding author(s).

\section*{Acknowledgements}

We are grateful to Claudia de Rham, H\'{e}ctor Gil-Mar\'{i}n, Olga Mena, Fergus Simpson, and Luca Visinelli for useful discussions and suggestions. S.V. is supported by the Isaac Newton Trust and the Kavli Foundation through a Newton-Kavli fellowship, and by a grant from the Foundation Blanceflor Boncompagni Ludovisi, n\'{e}e Bildt. S.V. acknowledges a College Research Associateship at Homerton College, University of Cambridge. D.F.M. acknowledges support from the Research Council of Norway and UNINETT Sigma2 -- the National Infrastructure for High Performance Computing and Data Storage in Norway. M.B. is supported by the grants ASI n.I/023/12/0, ASI-INAF n. 2018-23-HH.0, PRIN MIUR 2015 ``Cosmology and Fundamental Physics: illuminating the Dark Universe with Euclid'', and PRIN MIUR 2017 ``Combining Cosmic Microwave Background and Large Scale Structure data: an Integrated Approach for Addressing Fundamental Questions in Cosmology'' (2017YJYZAH).

\bsp
\label{lastpage}
\end{document}